\documentclass[%
reprint,
 amsmath,amssymb,
 aps,natbib,pre
]{revtex4-2}
\usepackage{geometry}
\usepackage{comment}
\usepackage{enumerate}
\usepackage{amsmath}
\usepackage{xcolor}
\usepackage{mathrsfs}
\usepackage{tabularx}
\usepackage{graphicx}
\usepackage{dcolumn}
\usepackage{appendix}
\usepackage{lipsum}
\usepackage{natbib}
\usepackage{bm}

\usepackage[mathlines]{lineno}
\usepackage{xcolor}
\usepackage{color}
\usepackage[colorlinks=true,linktoc=page,citecolor=blue,linkcolor=blue,]{hyperref}
\usepackage{graphicx}
\usepackage{subcaption}
\usepackage[section]{placeins}
\usepackage[export]{adjustbox}
\usepackage{float}

\begin{document}

\title{ Universal energy cascade in critically balanced homogeneous ferrofluid turbulence 
		} 
\author{Sukhdev Mouraya}

\author{Supratik Banerjee}
\homepage{Author to whom correspondence should be addressed: sbanerjee@iitk.ac.in}

\author{Nandita Pan}
\affiliation{
 Department of Physics, Indian Institute of Technology Kanpur, Uttar Pradesh, 208016 India
}%
%
\date{\today}
\begin{abstract}
In ferrofluids, the vorticity is balanced by the rate of particle rotation, which is known here as critical balance. The universal energy cascade is investigated for a stationary and non-stationary incompressible ferrofluid turbulent system using exact relations. The findings reveal that under moderate magnetic fields, kinetic and total energy cascades occur at identical rates. As the external magnetic field strength increases, the total energy cascade becomes non-stationary and differs from the kinetic one. However, the cascade's universal, scale-independent nature remains unaffected. The suppression of turbulence in ferrofluids is observable at strong magnetic fields.
\end{abstract}
\maketitle

\section{Introduction}

Ferrofluids are stable suspensions of nanosized ferromagnetic particles in a carrier fluid \cite{BAFinlayson1970, Rosensweig1997}. A wide range of disciplines, including biomedicine, fluid dynamics, robotics, microfluidics, optics, and nanotechnology, are covered by ferrofluids \cite{BAFinlayson1970, Schumacher2008, Shokrollahi2013, Koh2013, Jing2021, Chung2021}. The magnetic nanoparticles are mixed in the carrier fluid in such a way that it flows with the molecules of the carrier fluid and the whole combination behaves like a normal fluid with magnetic properties \cite{Shliomis1972, Zebib1996, Kaloni2004, Shliomis1994, Bacri1995, BAFinlayson2013}. Due to the presence of magnetic particles, ferrofluids respond to the magnetic field and form a chain-like structure which obstructs mobility \cite{Valentin2004, Dixit2020, Li2023}. Mouraya et. al. \cite{Mouraya2024} show that this chain-like structure can be reduced by generating turbulence in ferrofluid flow by strong forcing. Turbulent flows are highly irregular and chaotic. Turbulence is characterized by randomness in velocity through irregular fluctuations and vortices across various scales. In turbulence, energy is transferred from large scales (where it is input as forcing) to smaller scales (where it is dissipated as heat) through a process known as the energy cascade \cite{Monin1975, Karman1938, Frisch1995}. The energy cascade is a central concept in turbulence that describes how energy injected into a fluid system at large scales is transferred to progressively smaller scales until it is ultimately dissipated as heat and characterized by constant transfer rate $\varepsilon$. This process is crucial to understanding the behavior of turbulent flows and how they evolve over time.

The core of the energy cascade occurs in the inertial range, a range of intermediate scales where energy is passed from large eddies to smaller eddies without much loss to viscous dissipation. This transfer is driven by nonlinear interactions among the eddies. Larger eddies break down into smaller eddies, and this process repeats itself, moving energy through successively smaller scales \cite{Kolmogorov1941, Frisch1995, Argoul1989, Banerjee2014}. A key theory in turbulence, proposed by Kolmogorov, states that in this inertial range, the energy spectrum $E(k)$ follows a power-law distribution, specifically $ E(k) \sim k^{-5/3} $, where $k$ is the wavenumber corresponding to the size of the eddy \cite{Karman1938, Kolmogorov1941}. This theory is based on the assumption that the cascade process is universal and self-similar at these scales . In real space, the exact relation in turbulence theory refers to precise mathematical expressions derived directly from the governing equations of fluid motion, particularly for turbulence in homogeneous, isotropic flows \cite{Kolmogorov1941, Politano1998, politano1998dynamical, Galtier2008}. The universality in incompressible turbulence is studied using an exact law relating $\varepsilon$ to the statistical moments of the field variables \cite{Karman1938, Monin1975, Nandita2022}. These relations are crucial for understanding energy transfer, dissipation, and statistical properties in turbulence. One of the most famous exact relations is the Kolmogorov 4/5th law, which describes the energy cascade in three-dimensional turbulence. For incompressible HD and MHD turbulence, an alternate form of the exact relation including the second-order statistical moments is suggested \cite{Galtier2011, Banerjee2013, Banerjee2014, Banerjee2016, Banerjee2016b,  Banerjee2017, Banerjee2018, Nahuel2019}. This type of expression is especially interesting in situations where $\varepsilon$ cannot be expressed in terms of the divergence of the two-point fluctuations.

The study of energy cascade in ferrofluids turbulence is much more complex than neutral fluids as the interaction between ferrofluid and magnetic field leads to additional terms in the Navier-Stokes equation, which greatly modifies the corresponding turbulence. Majority of attention in ferrofluid research is directed towards investigating laminar flow regimes \cite{BAFinlayson1970, Shliomis1994, Bacri1995, Rosensweig1997, Felderhof1999, Kaloni2004, Papadopoulos2012, BAFinlayson2013, Buschmann2020}  and only a few studies are performed for turbulent ferrofluids \cite{Adrian1981, Schumacher2003, Schumacher2008, Schumacher2010, Schumacher2011, Altmeyer2015}. Schumacher \cite{Schumacher2008} for the first time studied the cascade of kinetic energy for a fully developed homogeneous ferrofluid turbulent flow in a steady magnetic field. However, kinetic energy is not an inviscid invariant of the incompressible ferrofluid flow. Rather, the total energy which is the sum of kinetic energy and work done due to magnetization is an inviscid invariant \cite{Mouraya2019, Mouraya2024}. More importantly, the ferrofluid particles always try to align themselves along the ambient magnetic field which causes constant leakage of energy at all scales from the fluid to the external field. The conservation of energy is only guaranteed when the flow is highly turbulent and the leakage is negligibly small. In fully developed turbulence, the inviscid invariants are expected to cascade from one scale to the other with a constant transfer rate ($\varepsilon$). Therefore, the total energy of the ferrofluid should exhibit a scale-independent transfer inside the inertial range of length scales. Such a scale-independent transfer signifies the presence of universality in ferrofluid turbulence. Hence, a detailed study on the total energy cascade has been important. 

The derivation of exact relations in ferrofluid is challenging due to the complex nature of the governing equations compared to other systems such as hydrodynamic (HD) and magnetohydrodynamic (MHD) fluids \cite{Karman1938, Politano1998, politano1998dynamical, Politano2003, Galtier2011, Monin1975, Antonia1997, Podesta2008, Augier2012, Nandita2022, Banerjee2011, Banerjee2013, Banerjee2014, Banerjee2016, Banerjee2017, Banerjee2020}. Despite the complexity, for the first time Mouraya and Banerjee \cite{Mouraya2019} derived the exact relation for an incompressible ferrofluid turbulence using the alternative approach which holds good for theoretical assumption. For numerical analysis or realistic assumption, Mouraya et. al. \cite{Mouraya2024} have derived the exact relation and numerical show that at high external field a non-stationary cascade occurs with constant cascade rate. In the course of our study in this paper we followed the paper \cite{Mouraya2024} and derived the exact relation for the situation where the magnetic particles rotates with the vorticity of the fluid and validated the universality of such exact relations through direct numerical simulations (DNSs).

The paper is organized as follows. In Sec.~\ref{sec:level2} the governing equations, conservation of total energy and simulations details for the ferrofluid system are described whereas Sec.~\ref{sec:level3} contains the derivation of the exact relation for negligible ferrofluid particle size. In Sec.~\ref{sec:level4}, we present the simulation parameters and numerical methods. In Sec.~\ref{sec:level5}, we present and discuss our results and findings. Finally, in Sec.~\ref{sec:level6}, we summarize and conclude.

{\section{\label{sec:level2}Theory and simulation details}}

{\subsection{Basic equations}}
The transport equations for incompressible ferrohydrodynamics consist of the linear momentum, internal angular momentum, and magnetization \citep{Schumacher2008, Mouraya2019, Mouraya2024} where density is kept constant and is given by
\begin{align}
 \left( \partial_t + \textbf{v} \cdot \boldsymbol{\nabla} \right) \textbf{v}  &=- \boldsymbol{\nabla} p  + \nu \nabla^2 \textbf{v} + \mu_0 (\textbf{M} \cdot \boldsymbol{\nabla}) \textbf{H} \nonumber \\
 &- \zeta \boldsymbol{\nabla} \times (  \boldsymbol{\Omega}-2\boldsymbol{\omega}) 
, \label{1} \\
 I \left( \partial_t + \textbf{v} \cdot\boldsymbol{\nabla} \right) \boldsymbol{\omega}  &=  \mu_0 {(\textbf{M} \times \textbf{H})} +\eta \nabla^2 \boldsymbol{\omega} + 2 \zeta (\boldsymbol{\Omega}-2\boldsymbol{\omega}), \label{2}	\\
(\partial_t + \textbf{v} \cdot \boldsymbol{\nabla}) \textbf{M} &=\boldsymbol{\omega} \times \textbf{M}-\frac{1}{\tau}(\textbf{M}-\textbf{M}_{eq}), \label{3} \\
\boldsymbol{\nabla} \cdot \textbf{v} &= 0, \hspace{0.2cm} \boldsymbol{\nabla} \times \textbf{H} = \textbf{0}, \label{4}
\end{align}
where $\textbf{v}$ is the velocity of the ferrofluid, $\boldsymbol{\Omega} = \boldsymbol{\nabla} \times \textbf{v}$ is the vorticity, $p$ is the fluid pressure, $\boldsymbol\omega$ is the ferrofluid particle spin rate, $\textbf{M}$ is the magnetization vector, $\textbf{H}$ is the magnetic field vector, $I$ is the  moment of inertia per unit mass for a ferrofluid particle, $\nu$ is the kinematic viscosity, $\zeta$ is the vortex viscosity,
$\eta$ is the spin viscosity and $\tau$ is the relaxation time.
The equilibrium magnetization is given by $\textbf{M}_{eq} = M_s L(\xi) {\textbf{H}}/{H}$, where 
$ L(\xi) = (\xi \coth(\xi)-{1})/{\xi} $ is the Langevin function with $\xi = {\mu_0 m H}/{k_B T} $ for a ferrofluid at temperature $T$. The parameters $M_s$ and $m$ are the magnitudes of saturation magnetization and magnetic moment of a single ferrofluid particle respectively. For small values of $\xi$, one can write $
\textbf{M}_{eq} = \chi \textbf{H}$, where $\chi$ is magnetic susceptibility. Due to incompressibility, $\textbf{v}$ is divergence-less and in the absence of any free current $\textbf{H}$ is irrotational. Since, the magnetic flux (\textbf{B}) is divergence-free, the evolution of the magnetic field can directly be obtained from that of the magnetization as $\textbf{B} = \mu_0 (\textbf{H} + \textbf{M})$ and hence one can write 
\begin{equation}
    \boldsymbol{\nabla} \cdot \textbf{M} = - \boldsymbol{\nabla} \cdot \textbf{H}. \label{5}
\end{equation}

For a realistic approach or numerical analysis one has to consider that the specific moment of inertia $ I (\sim 10^{-16} m^2)$ and the spin viscosity $\eta~(\sim 10^{-15} kg \ m \ s^{-1})$ are very small and can be neglected with respect to the other terms in the evolution equation of internal angular momentum \citep{Schumacher2008, Mouraya2024}. Hence, the angular momentum Eq.~\eqref{2} reduces to
\begin{equation}
    \boldsymbol{\omega} = \frac{\boldsymbol{\Omega}}{2} +\frac{\mu_0}{4 \zeta} (\textbf{M} \times \textbf{H}). \label{5}
\end{equation}
Using Eq. \eqref{5} in Eqs. \eqref{1} and \eqref{3}, one obtains
\begin{align}
 ( \partial_t + \bm{v} \cdot \boldsymbol{\nabla}) \bm{v} &=- \boldsymbol{\nabla} p + \nu \nabla^2 \bm{v} + \mu_0 (\textbf{M} \cdot \boldsymbol{\nabla}) \textbf{H} \nonumber \\ & + \frac{\mu_0}{2} \boldsymbol{\nabla} \times (\textbf{M} \times \textbf{H})
, \label{6}\\	
(\partial_t + \bm{v} \cdot \boldsymbol{\nabla}) \textbf{M} &= \frac{1}{2} (\boldsymbol{\Omega} \times \textbf{M}) + \frac{\mu_0}{4 \zeta} (\textbf{M} \times \textbf{H}) \times \textbf{M} \nonumber \\ & - \frac{1}{\tau}(\textbf{M}-\chi \textbf{H}), \label{7}
\end{align} 

For the realistic flow assumption, the ferrofluid turbulence is governed by the equations of motion mentioned above.

\subsection{Conservation of total energy}

The total energy is determined by the kinetic energy of the fluid and the internal energy resulting from the work performed by the ferromagnetic particles in response to the external magnetic field. Using  Eqs.~\eqref{5} and \eqref{6}, the kinetic energy evolution equation can be written as

\begin{align}
   & \partial_t \left(\frac{v^2}{2}\right) +\boldsymbol{\nabla} \cdot \left(\frac{v^2}{2} \textbf{v}\right) =-\boldsymbol{\nabla} \cdot (p \textbf{v} )+ \mu \textbf{v} \cdot \nabla^2 \textbf{v} \nonumber \\ & + \frac{\mu_0}{2} \boldsymbol{\nabla} \cdot ( (\textbf{M} \times \textbf{H}) \times \textbf{v} ) + \frac{\mu_0}{2} (\textbf{M} \times \textbf{H}) \cdot \boldsymbol{\Omega} \nonumber \\ &
   + \mu_0 \textbf{v} \cdot (\textbf{M} \cdot {\boldsymbol{\nabla}}) \textbf{H} + \textbf{v} \cdot \textbf{f}_v,
\end{align}
and also from Eqs. \eqref{5} and \eqref{7}, we get

\begin{align}	
& \textbf{H} \cdot \partial_t \textbf{M}  = - \textbf{H} \cdot ((\textbf{v} \cdot \boldsymbol{\nabla}) \textbf{M}) + \frac{1}{2} \textbf{H} \cdot(\boldsymbol{\Omega} \times \textbf{M})  \nonumber \\ & + \frac{\mu_0}{4 \zeta} \textbf{H} \cdot ((\textbf{M} \times \textbf{H}) \times \textbf{M}) -\frac{1}{\tau} \textbf{H} \cdot (\textbf{M}-\chi \textbf{H}) \nonumber \\ 
& = - \textbf{H} \cdot ((\textbf{v} \cdot \boldsymbol{\nabla}) \textbf{M}) + \frac{1}{2} \textbf{H} \cdot(\boldsymbol{\Omega} \times \textbf{M}) + \frac{\mu_0}{4 \zeta} (\textbf{M} \times \textbf{H})^2 \nonumber \\ & - \frac{1}{\tau} \textbf{H} \cdot (\textbf{M}-\chi \textbf{H})\nonumber \\ 
& = - \textbf{H} \cdot ((\textbf{v} \cdot \boldsymbol{\nabla}) \textbf{M}) + \frac{1}{2} \textbf{H} \cdot(\boldsymbol{\Omega} \times \textbf{M}) + \frac{\zeta}{\mu_0} ( 2 \boldsymbol{\omega} - \boldsymbol{\Omega} )^2 \nonumber \\ & -\frac{1}{\tau} \textbf{H} \cdot (\textbf{M}-\chi \textbf{H})
\end{align}

The evolution equation of the total energy is given by
\begin{align}
\partial_t E =& \partial_t \int \frac{v^2}{2} d\tau - \int \mu_0 \textbf{H} \cdot \partial_t \textbf{M} d\tau \nonumber \\ =& - \int \boldsymbol{\nabla} \cdot \left[ \left( \frac{v^2}{2} + p + \mu_0 \textbf{M} \cdot \textbf{H} \right) \textbf{v} \right. \nonumber \\ & \left.  - \frac{\mu_0}{2} (\textbf{M} \times \textbf{H}) \times \textbf{v} \right] d \tau - \int \zeta ( 2 \boldsymbol{\omega} - \boldsymbol{\Omega} )^2 d\tau \nonumber \\ & + \int \left( \mu \textbf{v} \cdot \nabla^2 \textbf{v} + \frac{\mu_0}{\tau} \textbf{H} \cdot (\textbf{M} - \chi \textbf{H}) \right) d \tau. \label{10}
\end{align}
The first term in the RHS of Eq.~\eqref{10} vanishes by the use of the Gauss divergence theorem with periodic or vanishing boundary conditions. In the inviscid limit, the term $\int \mu \textbf{v} \cdot \nabla^2 \textbf{v} d \tau$  can be ignored. Also, for turbulent flows the term $ \frac{\mu_0}{\tau} \int  \textbf{H} \cdot (\textbf{M} - \chi \textbf{H}) d \tau$ vanishes \cite{Mouraya2019}. Hence, the evolution equation for total energy is
\begin{align}
 \partial_t \int \frac{v^2}{2} d\tau - \int \mu_0 \textbf{H} \cdot \partial_t \textbf{M} d\tau &= - \int \zeta ( 2 \boldsymbol{\omega} - \boldsymbol{\Omega} )^2 d \tau, \label{13}
\end{align}
Here $ \zeta \int (2 \boldsymbol{\omega} - \boldsymbol{\Omega} )^2 d \tau = \zeta \int (4 \omega^2 +  \Omega^2 - 4 \boldsymbol{\omega} \cdot \boldsymbol{\Omega}) d \tau = 4 \zeta \int \boldsymbol{\omega} \cdot(\boldsymbol{\omega} - \boldsymbol{\Omega}) d \tau$ as $ \zeta \int \Omega^2 d \tau \rightarrow 0$ in the inviscid limit. Now, energy conservation can be shown either by considering $\zeta \int (2 \boldsymbol{\omega} - \boldsymbol{\Omega})^2 d \tau$ negligibly small or by taking $\boldsymbol{\omega} = \boldsymbol{\Omega}$. Note that $2\boldsymbol{\omega} = \boldsymbol{\Omega}$ would also lead to energy conservation. However, one cannot accept this condition since then Eq. \eqref{5} would lead to $\textbf{\textbf{M}} || \textbf{H}$, which is the relaxed state of the system. For negligible $\zeta (2 \boldsymbol{\omega} - \boldsymbol{\Omega})^2$, the energy conservation and exact relation associate to this is shown in paper \cite{Mouraya2024}. For $\boldsymbol{\Omega} = \boldsymbol{\omega}$, refers to as a critically balanced flow, the evolution equation for total energy is given by
\begin{align}
    \partial_t E &= \partial_t \int \frac{v^2}{2} d \tau - \int \mu_0 \textbf{H} \cdot \partial_t \textbf{M}  \ d \tau = 0 \label{energy_conservation}
\end{align}
and the dissipation term is as follows
\begin{align}
    d &= \int \left( (\mu + \zeta)  \textbf{v} \cdot \nabla^2 \textbf{v} + \frac{\mu_0}{\tau} \textbf{H} \cdot (\textbf{M}-\textbf{M}_0) \right) d \tau.
\end{align}
Here, it can be observed that the effective viscosity of the ferrofluid decreases because the alignment reduces inter-particle collisions and rotational resistance which influences the dissipation rate. Now, the evolution equation for momentum and magnetization for the critical flow is given by
\begin{align}
    \partial_t \textbf{v} &= - (\textbf{v} \cdot \boldsymbol{\nabla}) \textbf{v} - \boldsymbol{\nabla} p + (\mu + \zeta) \nabla^2 \textbf{v} + \mu_0 (\textbf{M} \cdot \boldsymbol{\nabla}) \textbf{H} 
    \nonumber \\ & + \mu_0 \boldsymbol{\nabla} \times ( \textbf{M} \times \textbf{H}), \label{14} \\	
    \partial_t \textbf{M} &= -(\textbf{v} \cdot \boldsymbol{\nabla}) \textbf{M} + \boldsymbol{\Omega} \times \textbf{M} - \frac{1}{\tau}(\textbf{M} - \textbf{M}_0), \label{15}
\end{align}
which constitute the revised set of governing equations for the ferrofluid turbulence for the realistic assumption of flow properties when particle glides with the vorticity of the flow (critically balanced flow). Also, the magnetization dynamics are significantly affected due to the alignment of particle motion with the flow's rotational behavior. Again, the total energy is inviscid invariant in the ferrofluid flow. Knowledge about the cascading of energy from one length scale to another is achievable through comprehending the length scales at two distinct points. Hence, an evolution equation of the two-point energy correlator and the exact relation associated with it can be obtained. 
For the sake of numerical implementation, it is necessary to rewrite the above set of equations in terms of dimensionless starred variables as
\begin{align}
    \partial_{t^*} \textbf{v}^* &= - (\textbf{v}^* \cdot \boldsymbol{\nabla}^*) \textbf{v}^* - \boldsymbol{\nabla}^* p^* + \frac{1.55}{Re} \nabla^{*2} \textbf{v}^* \nonumber \\ & + (\textbf{M}^* \cdot \boldsymbol{\nabla}^*) \textbf{H}^* 
    + \boldsymbol{\nabla}^* \times ( \textbf{M}^* \times \textbf{H}^*) + \textbf{f}^{*}_{v}, \label{15} \\	
    \partial_{t^*} \textbf{M}^* &= -(\textbf{v}^* \cdot \boldsymbol{\nabla}^*) \textbf{M}^* + \boldsymbol{\Omega}^* \times \textbf{M}^* - \frac{1}{\Gamma}(\textbf{M}^* - \chi \textbf{H}^*), \label{16}
\end{align}
where $\zeta = 0.55\mu$ (as used in \citep{Schumacher2008}), $ \textbf{v} = v_{rms} \textbf{v}^*$, $\textbf{M} = \sqrt{\frac{1}{\mu_0}} v_{rms} \textbf{M}^*$, $\textbf{H} = \sqrt{\frac{1}{\mu_0}} v_{rms} \textbf{H}^*$, $t = \frac{l}{v_{rms}} t^*$, $p = v_{rms}^2 p^*$, $\Gamma = \frac{v_{rms}}{l_o} \tau$, $Re = \frac{v_{rms} l}{\mu}$ is the large-scale Reynolds number, with $l_0$ representing the box size and $v_{rms}$ the root mean square velocity. In order to simplify the notations, we shall omit the stars from the dimensionless variables hereinafter and the system is propelled by a large-scale forcing in the momentum equation to produce a persistent turbulent flow.

\subsection{Simulation details}
To set up the initial conditions, we initially developed the turbulent flow using the pure Navier-Stokes equation of a normal fluid. Then, we use this developed velocity as the initial velocity to study the ferrofluid flow. To solve the magnetization field we use a random initial condition. A uniform time-independent external magnetic field of strength H0 = 0.1 and H0 = 1.0 is applied along the z direction. Finally, to sustain the flow, energy is injected by forcing the momentum evolution equation with a large-scale Taylor-Green forcing, given by 
$\bm{f}_v \equiv f_0[sin(k_ox)cos(k_oy)cos(k_oz), -cos(k_ox)sin(k_oy) \\
cos(k_oz),0]$, where $k_o = 2$ is the energy-injection scale and $f_0$ is the forcing amplitude. For time evolution, we use the fourth-order Runge-Kutta (RK4) method. The system evolves until it reaches a statistically stationary state. Once the stationary state is achieved we can use the data to numerically verify the exact law.

\section{\label{sec:level3} Derivation of Exact relation}

The total energy is an inviscid invariant of flow. Hence, an exact law can be obtained for the critically balanced flow. The evolution equation of the correlation function by using Eqs. \eqref{15} and \eqref{16} is

\begin{align}
\frac{\partial \mathcal{R}}{\partial t} &= \frac{1}{2}  \left< \frac{\partial \textbf{v}}{\partial t}  \cdot \textbf{v}' +  \textbf{v} \cdot \frac{\partial \textbf{v}'}{\partial t} - \left( \textbf{H} \cdot \frac{\partial \textbf{M}'}{\partial t} + \textbf{H}' \cdot \frac{\partial \textbf{M}}{\partial t} \right) \right> \nonumber \\ 
&= \frac{1}{2}\left< (\textbf{v} \times \boldsymbol{\Omega} ) \cdot \textbf{v}'+(\textbf{v}' \times \boldsymbol{\Omega}') \cdot \textbf{v} + \textbf{v} \cdot (\textbf{M}' \cdot \boldsymbol{\nabla}') \textbf{H}' \right. \nonumber \\ & \left. + \textbf{v}' \cdot (\textbf{M} \cdot \boldsymbol{\nabla}) \textbf{H} + \textbf{H} \cdot (\textbf{v}' \cdot \boldsymbol{\nabla}') \textbf{M}' + \textbf{H}' \cdot (\textbf{v} \cdot \boldsymbol{\nabla}) \textbf{M} \right. \nonumber \\
& \left. - \textbf{H}' \cdot (\boldsymbol{\Omega} \times \textbf{M}) - \textbf{H} \cdot (\boldsymbol{\Omega}' \times \textbf{M}') \right.
\nonumber \\
& \left. + \textbf{v}' \cdot (\boldsymbol{\nabla} \times (\textbf{M} \times \textbf{H})) + \textbf{v} \cdot (\boldsymbol{\nabla}' \times (\textbf{M}' \times \textbf{H}')) \right.
\nonumber \\
& \left. - \textbf{v}  \cdot \boldsymbol{\nabla}' \left(p' + \frac{v^{'2}}{2} \right) - \textbf{v}'  \cdot \boldsymbol{\nabla} \left( p + \frac{v^{2}}{2} \right) \right> \nonumber \\ & + D + F, \label{c4eq72}
\end{align}
where $D$ and $F$ represent dissipation and forcing terms respectively, and are given by:
\begin{align}
&D = \frac{1}{2}  \left< \frac{1.55}{Re} \textbf{v}' \cdot \nabla^2 \textbf{v} + \frac{1.55}{Re} \textbf{v} \cdot \nabla'^2 \textbf{v}' \right. \nonumber \\ & \left. + \frac{1}{\Gamma} \textbf{H}' \cdot (\textbf{M}-\chi \textbf{H}) + \frac{1}{\Gamma} \textbf{H} \cdot (\textbf{M}'-\chi \textbf{H}')\right>,\\
&F =\frac{1}{2} \left< \textbf{v} \cdot \textbf{f}'_v + \textbf{v}' \cdot \textbf{f}_v \right> + \text{mean terms}.
\end{align}
Using statistical homogeneity, one can prove that
\begin{align}
    & (i) \left<(\textbf{v} \times \boldsymbol\Omega ) \cdot \textbf{v}'+(\textbf{v}' \times \boldsymbol\Omega') \cdot \textbf{v} \right> =- \left< \delta(\textbf{v} \times \boldsymbol\Omega)\cdot \delta \textbf{v} \right> \label{19}\\
    & (ii) \left< \textbf{v} \cdot (\textbf{M}' \cdot \boldsymbol{\nabla}') \textbf{H}' + \textbf{v}'\cdot (\textbf{M} \cdot \boldsymbol{\nabla}) \textbf{H} + \textbf{H} \cdot (\textbf{v}' \cdot \boldsymbol{\nabla}') \textbf{M}' \right. \nonumber \\ & \left. 
    + \textbf{H}' \cdot (\textbf{v} \cdot \boldsymbol{\nabla}) \textbf{M} \right> = \left< - \delta \textbf{v} \cdot \delta((\textbf{M} \cdot \boldsymbol{\nabla}) \textbf{H}) \right. \nonumber \\ & \left. - \delta \textbf{H} \cdot \delta((\textbf{v} \cdot \boldsymbol{\nabla}) \textbf{M}) \right> \label{20}\\ 
     & (iii)  \left< \textbf{H}' \cdot (\boldsymbol{\Omega} \times \textbf{M}) + \textbf{H} \cdot (\boldsymbol{\Omega}' \times \textbf{M}') \right. \nonumber \\ & \left. - \textbf{v}' \cdot (\boldsymbol{\nabla} \times (\textbf{M} \times \textbf{H})) - \textbf{v} \cdot (\boldsymbol{\nabla}' \times (\textbf{M}' \times \textbf{H}')) \right> \nonumber \\
     & = \left< \delta \bm{\Omega} \cdot \delta ( (\textbf{M} \times \textbf{H})) - \delta \textbf{H} \cdot \delta (\boldsymbol{\Omega} \times \textbf{M}) \right>. \label{21}
\end{align}

Using incompressible and statistical homogeneity and combining the equations $\eqref{19}$, $\eqref{20}$, $\eqref{21}$ and putting in equation $\eqref{c4eq72}$, we get
\begin{align}
\frac{\partial \mathcal{R}}{\partial t} 
&= \frac{1}{2}\left<- \delta(\textbf{v} \times \boldsymbol{\Omega} ) \cdot \delta\textbf{v} - [ \delta\textbf{v} \cdot \delta (\textbf{M} \cdot \boldsymbol{\nabla} \textbf{H}) \right. \nonumber \\ & \left. - \delta \textbf{H} \cdot \delta(\textbf{v} \cdot \boldsymbol{\nabla} \textbf{M}) - \delta \bm{\Omega} \cdot \delta (\textbf{M} \times \textbf{H}) \right. \nonumber \\ 
& \left. + \delta \textbf{H} \cdot \delta (\boldsymbol{\Omega} \times \textbf{M}) ] \right> + D + F.
\end{align}
Now, decomposing the total magnetic field as $\textbf{H} = \textbf{H}_0 + \textbf{\~H} = H_0 \hat{z} + \textbf{\~H}$, we obtain
\begin{align}
\frac{\partial \mathcal{R}}{\partial t}
&= \frac{1}{2}\left<- \delta(\textbf{v} \times \boldsymbol{\Omega} ) \cdot \delta\textbf{v} - [ \delta\textbf{v} \cdot \delta (\textbf{M} \cdot \boldsymbol{\nabla} \textbf{\~H}) \right. \nonumber \\ & \left. - \delta \textbf{\~H} \cdot \delta(\textbf{v} \cdot \boldsymbol{\nabla} \textbf{M}) - \delta \bm{\Omega} \cdot \delta (\textbf{M} \times \textbf{\~H})\right. \nonumber \\ 
& \left. + \delta \textbf{\~H} \cdot \delta (\boldsymbol{\Omega} \times \textbf{M}) + \textbf{H}_0 \cdot (\delta \boldsymbol{\Omega} \times \delta \textbf{M})] \right> + D + F.
\end{align}

Again, assuming statistical stationarity and inside the inertial zone, one can derive the following exact relation
\begin{align}
A_c(\boldsymbol{\ell}) = A_{c_1}(\boldsymbol{\ell}) + A_{c_2}(\boldsymbol{\ell}) = 2 \varepsilon_c \label{eq27}, 
\end{align} 
where $\varepsilon_c = F_c \approx \left< \textbf{v} \cdot \bm{f}_v + \frac{1}{\Gamma} \textbf{H}_0 \cdot (\textbf{M} - \chi \textbf{H}) \right> = \varepsilon_{inj} - \varepsilon_{c_{H_0}} $  is the mean energy injection rate.

The two-point dissipative terms can be written as
\begin{align}
    D_c &= \frac{1}{2}\left< \frac{1.55}{Re} \textbf{v} \cdot \nabla'^2 \textbf{v}' + \frac{1.55}{Re} \textbf{v}' \cdot \nabla^2 \textbf{v} \right. \nonumber \\ & \left. + \frac{1}{\Gamma} \textbf{\~H}' \cdot (\textbf{M} - \chi \textbf{H}) + \frac{1}{\Gamma} \textbf{\~H} \cdot (\textbf{M}' - \chi \textbf{H}') \right>, \label{27}
\end{align}
and the forcing terms $F$ are
\begin{align}
    F_c &= \frac{1}{2} \left< \textbf{v}' \cdot \bm{f}_v + \textbf{v} \cdot \bm{f}'_v \right \rangle + \left<  \frac{1}{\Gamma} \textbf{H}_0 \cdot (\textbf{M} - \chi \textbf{H}) \right>. \label{28}
\end{align}
and

\begin{align}
    A_{c_1} (\bm{\ell}) &= \left< \delta \textbf{v} \cdot \left[ \delta(\textbf{v} \times \boldsymbol\Omega)  +  \delta((\textbf{M} \cdot \boldsymbol{\nabla}) \textbf{\~H}) \right] \right. \nonumber \\ & \left. + \delta \textbf{\~H} \cdot \left[ \delta ((\textbf{v} \cdot \boldsymbol{\nabla}) \textbf{M}) -  \delta (\boldsymbol{\Omega} \times \textbf{M}) \right] \right. \nonumber \\ & \left. + \delta \boldsymbol{\Omega} \cdot  \delta (\textbf{M} \times \textbf{\~H}) \right> , \label{29}\\
 A_{c_2} (\bm{\ell}) &= - \left<  \textbf{H}_0\cdot (\delta \boldsymbol{\Omega} \times \delta \textbf{M}) \right>, \label{30}
\end{align}

It provides the energy cascade rate for turbulence in ferrofluids with negligible ferrofluid particles and the rotation of ferrofluid particles matches the vorticity of the fluid. Whereas $A_{c_1}(\ell)$ denotes the part of $\varepsilon_c$ coming exclusively from the fluctuations, $A_{c_2}(\ell)$ gives the contribution of the external magnetic field $H_0$ to $\varepsilon_c$. As is evident from Eq.~\eqref{30}, the mean-field effect is present in the exact relation. As shown from Eq.~\eqref{30} and \eqref{28}, the external magnetic field (${H}_0$) not only actively contributes to the inertial range energy transfer but also modifies the input energy injection ($\varepsilon_c$). Next, we numerically study whether ferrofluid turbulence exhibits an inertial range energy cascade for the external fields ${H}_0 = 0.1$, ${H}_0 = 0.5$, and ${H}_0 = 1.0$.

\section{\label{sec:level4} Numerical simulation}
The field variables (velocity, magnetic field, and magnetization) are obtained using DNS over a long enough time span to reach a statistically stationary condition and preserve the field variables at various time steps. Two points $\textbf{x}$ and $\textbf{x}+\boldsymbol{\ell}$ are chosen over the physical space in order to calculate the two-point fluctuation or correlation function. For the numerical implementation of exact law, we average over the physical space over the increment vector $\boldsymbol{\ell}$ for all the possible pairs of points as 
\begin{equation}
   A(\boldsymbol{\ell}) = \sum_{\textbf{x}}\frac{[\textbf{a}(\textbf{x}+\boldsymbol{\ell} - \textbf{a}(\textbf{x}))] \cdot [\textbf{b}(\textbf{x}+\boldsymbol{\ell}) - \textbf{b}(\textbf{x})]}{N_xN_yN_z} ,
\end{equation}
where ($N_x, N_y, N_z$) are the dimensions of the grid, \textbf{a} and \textbf{b} are field variables. Assuming statistical homogeneity and isotropy, exact scaling laws are obtained. The exact law is an average over the two-point increments of the field variables.  Consequently, in order to explore every direction in space, we require a whole range of increments. We vary the increment vector over 73 directions spanned by the base vectors {($1,0,0$), ($1,1,0$), ($1,1,1$), ($2,1,0$), ($2,1,1$) ($2,2,1$), ($3,1,0$), ($3,1,1$)} \cite{Nahuel2019, Ferrandthesis}. This is done by using Taylor's method \cite{Taylor2003} to calculate the exact law assuming isotropy and homogeneity. As a result, $\ A (\ell) = \sum A (\bm{\ell})/73$ is the energy cascade rate form. The set is cleared of any vectors that are a positive or negative multiple of another vector. This method allows us to compute the exact law at the grid point along the direction of the specified base vector. After that, we interpolate along the direction of the increment vector.
\subsection{Simulation parameters}

The simulation parameters corresponding to each run are summarised in Table \ref{tab3}.

\begin{table*}[hbt]
\caption{\textbf{Simulation parameters}. $M_0$ is the mean magnetization, $L$ is the integral length scale, $\lambda$ is the Taylor length scale, $\eta$ is the Kolmogorov length scale, $U_{rms}$ is the rms (root mean square) velocity,  $Re_L $ is the integral-scale Reynolds number, $Re_{\lambda}$ is the Taylor-scale Reynolds number, the subscript a,  b, and c represent runs with $H_0 = 0.1$, 0.5 and 1.0 respectively.}
\label{tab3}
\begin{minipage}{1.0\textwidth}
{ \small
\begin{tabular}{c c c c c c c c c c c c}
\hline\hline
Run & \hspace{0.15cm} $N$ \hspace{0.15cm} & \hspace{0.15cm} $M_{0}$ \hspace{0.15cm} & \hspace{0.15cm} $U_{rms}$ \hspace{0.15cm} & \hspace{0.5cm} $L$ \hspace{0.5cm} & \hspace{0.5cm} $\lambda$ \hspace{0.5cm} & \hspace{0.5cm} $\eta$ \hspace{0.5cm} & \hspace{0.15cm} $Re_{\lambda}$ \hspace{0.15cm} & \hspace{0.15cm} $Re_L$ \hspace{0.15cm} & \hspace{0.15cm} $k_{max}/k_{\eta}$ \hspace{0.15cm} & \hspace{0.15cm} $L/\lambda$ \hspace{0.15cm} & \hspace{0.15cm} $L/\eta$ \hspace{0.15cm} \\
\hline
1a & 128  & 0.01 &  0.7  &  0.67 & 0.408  & 0.212 &  60  & 117 & 1.439 & 1.64 & 3.16\\
2a & 256  & 0.01 & 0.751 & 0.584 & 0.303  & 0.125 & 91  & 219 & 1.697 & 1.93 & 4.67\\
3a & 512  & 0.009 & 0.799 & 0.518 & 0.192 & 0.057 & 158 & 592 & 1.548 & 2.7 & 9.1\\
\hline
1b & 128  & 0.053 & 0.667 & 0.726 & 0.491 & 0.234 &  72  & 121 & 1.589 & 1.48 & 3.1\\
2b & 256  & 0.052 & 0.713 & 0.703 & 0.433 & 0.152 & 131 & 251 & 2.064 & 1.62 & 4.62\\
3b & 512  & 0.051 & 0.763 & 0.662 & 0.346 & 0.078 & 304 & 722 & 2.118 & 1.91 & 8.49\\
\hline
1c & 128  & 0.11 & 0.637 & 0.841 & 0.622  & 0.265 & 90 & 134 & 1.799 & 1.35 & 3.17\\
2c & 256  & 0.11 & 0.668 & 0.879 & 0.587  & 0.179 & 172 & 294 & 2.431 & 1.5 & 4.91\\
3c & 512  & 0.11 & 0.695 & 0.851 & 0.541  & 0.101 & 462 & 846 & 2.716 & 1.57 & 8.42
\\
\hline \hline
\end{tabular} }
\end{minipage}
\end{table*}

\subsection{Steady state and Statistical dynamics of field variables for ${H}_0 = 0.1$}

\begin{figure*}[hbt!]
\centering
\includegraphics[width=1.0\linewidth]{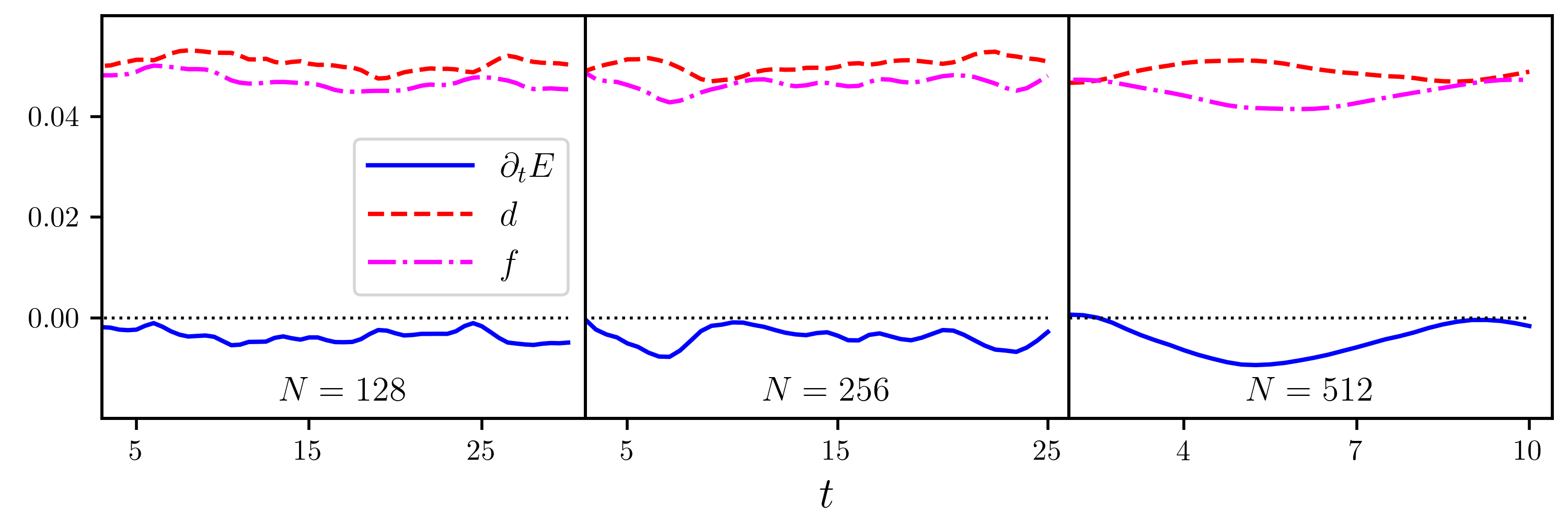}
\caption{Time evolution of the rate of total energy ($E$), dissipation ($d$) and injection ($f$) for grid size $256^3$ and $512^3$ Runs of Table \ref{tab3} 1a, 2a and 3a.}
\label{c5fig:10}
\end{figure*}

 \begin{figure*}[hbt!]
     \centering
    \includegraphics[width=0.95\linewidth]{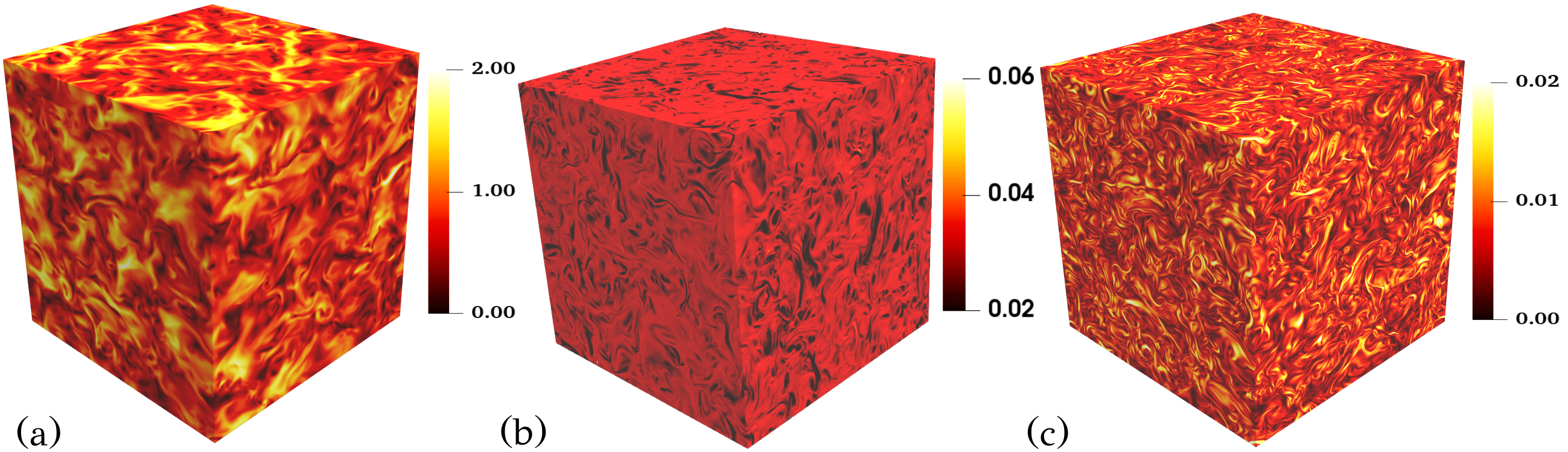}
         \caption{Snapshot of the Modulus of (a) velocity (b) magnetization and (c) magnetic field for $\textbf{H}_0 = 0.1$ (Table \ref{tab3}).}
         \label{c5fig:11}
\end{figure*}

\begin{figure*}[hbt!]
\centering
\includegraphics[width=1.0\linewidth]{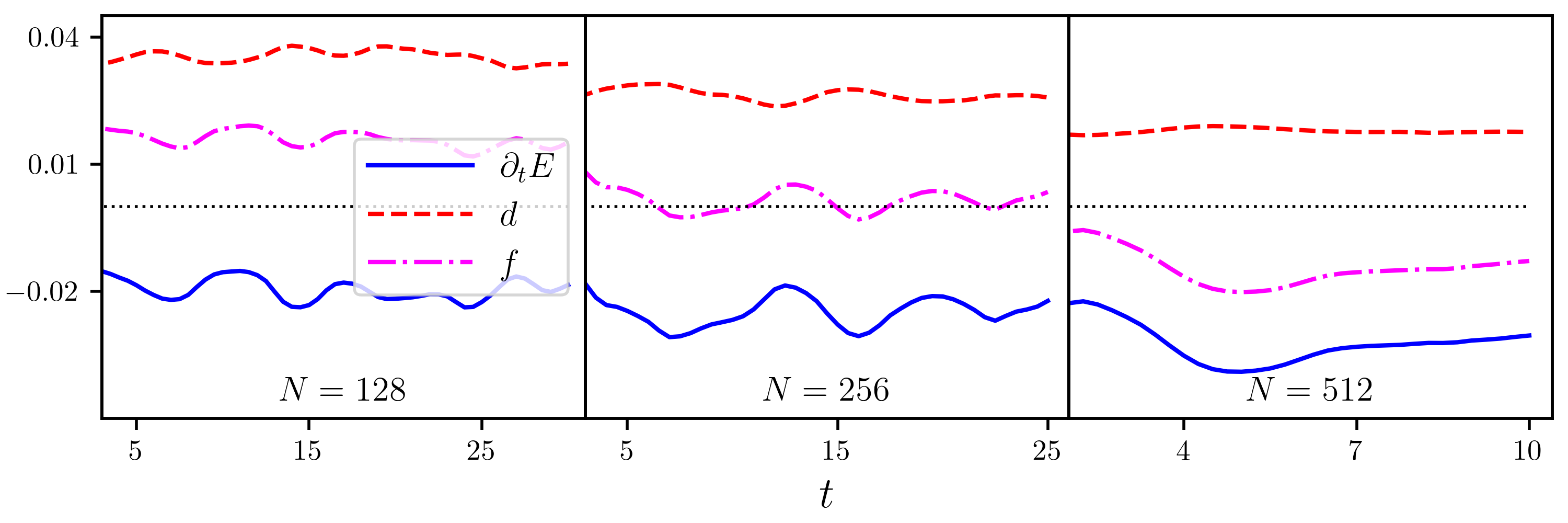}
\caption{Time evolution of the rate of total energy ($E$), dissipation ($d$) and injection ($f$) for grid size $128^3$, $512^3$ Runs of Table \ref{tab3} for $\textbf{H}_0 = 0.5$.}
\label{c5fig:12}
\end{figure*}

 \begin{figure*}[hbt!]
     \centering
    \includegraphics[width=0.95\linewidth]{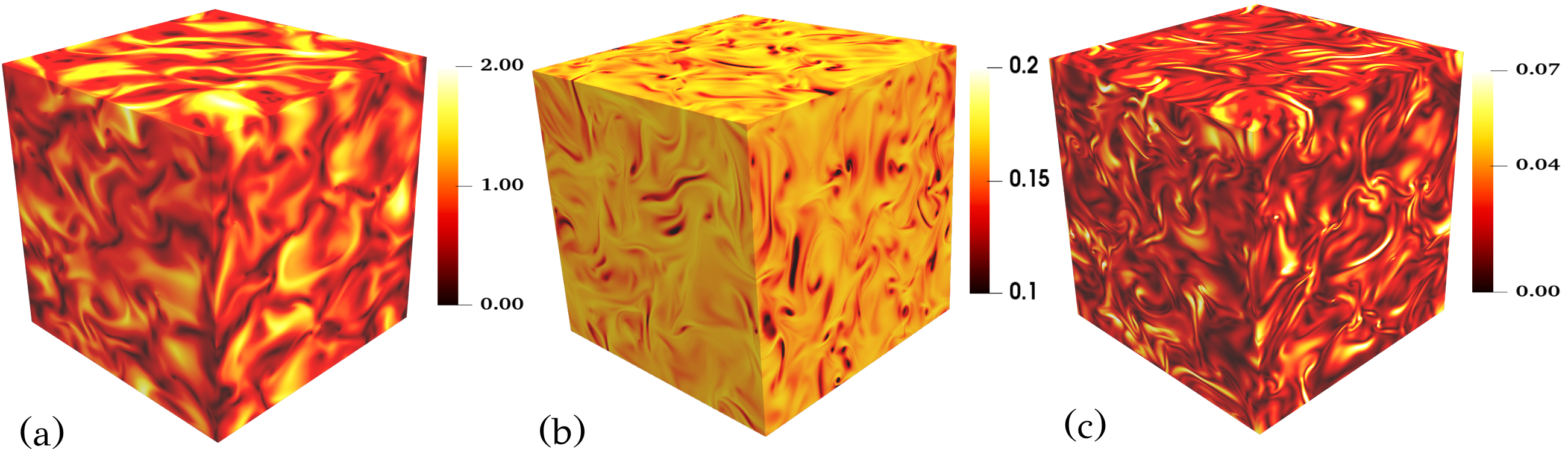}
         \caption{Snapshot of the Modulus of (a) velocity (b) magnetization and (c) magnetic field for Runs at $\textbf{H}_0 = 0.5$ (Table \ref{tab3}).}
         \label{c5fig:13}
\end{figure*}

As can be seen in Fig. \ref{c5fig:10}, the rate of total energy ($\partial_t E$) is very close to zero (horizontal dotted black line), indicating that the total energy is conserved. The average dissipation rate ($d$) is nearly equal to the average injection rate ($f$), which can be seen in Fig. \ref{c5fig:10}, and this causes the total energy to reach a statically stationary state. In the following part, every statistic is completed for the statistical steady state. 

Fig.~\ref{c5fig:11} shows that at the field $H_0 = 0.1$, the small-scale structures are present in all three field variables (velocity, magnetization and magnetic field). As a result, the energy cascade rate can be seen.

\subsection{Steady state and Statistical dynamics of field variables for ${H}_0 = 0.5$}

Now, we can observe the effect at high-field at the eternal field ${H}_0 = 0.5$.
Fig. \ref{c5fig:12} shows the plot of the total energy, dissipation rate and forcing rate. Rather than looking for a stationary state, we have collected data for a state where the average energy is changing at a steady rate over time using the same forcing as for $H_0 = 0.1$. Fig.~\ref{c5fig:13} shows that at the high field $H_0 = 0.5$, the small-scale structures are still present in all three field variables (velocity, magnetization and magnetic field) but in comparison to $H_0 = 0.1$ the structures are reduced. Hence there is less turbulence for $H_0 = 0.5$. As a result, a reduced energy cascade rate can be seen.

\subsection{Steady state and Statistical dynamics of field variables for ${H}_0 = 1.0$}

\begin{figure*}[hbt!]
\centering
\includegraphics[width=1.0\linewidth]{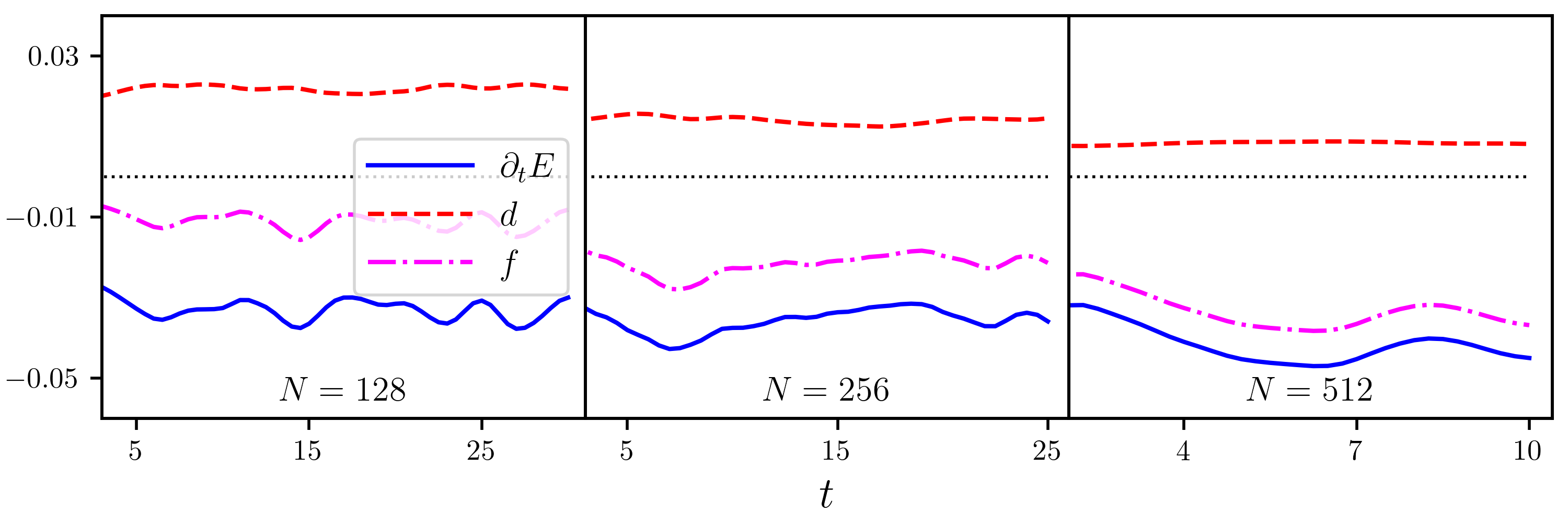}
\caption{Time evolution of the rate of total energy ($E$), dissipation ($d$) and injection ($f$) for grid size $128^3$, $512^3$ Runs of Table \ref{tab3} for $\textbf{H}_0 = 1.0$.}
\label{c5fig:14}
\end{figure*}

\begin{figure*}[hbt!]
     \centering
    \includegraphics[width=0.95\linewidth]{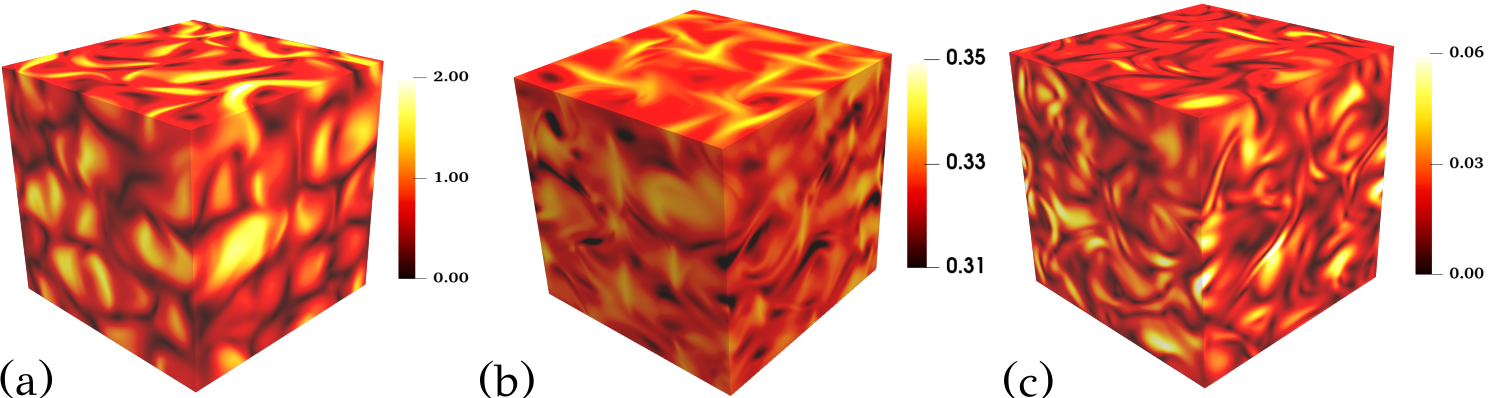}
         \caption{Snapshot of the Modulus of (a) velocity (b) magnetization and (c) magnetic field for Run $\textbf{H}_0 = 1.0$ (Table \ref{tab3}).}
         \label{c5fig:15}
\end{figure*}

In order to observe the effect at high-field, the field is now increased to ${H}_0 = 1.0$. Plots of the total energy, dissipation, and forcing rate are displayed in Fig. \ref{c5fig:14}. With a consistent change in average energy, we can observe the non-stationary condition in this case. Fig. \ref{c5fig:15} shows that at the high field $H_0 = 1.0$, only the large-scale structures are there in all three field variables. Hence, there is no turbulence or very weak turbulence for $H_0 = 1.0$ which means there is suppression of turbulence at large external field. Consequently, there is no inertial range or very small inertial range. As a result, there is no energy cascade rate. One can observe from Table \ref{tab3} that all the runs are well resolved as the ratio of the maximum wavenumber $k_{max} = N/3$ to the Kolmogorov wave number $k_{\eta} = 2 \pi / {\eta}$ where $\eta =  (\nu^3/\varepsilon_c)^{1/4}$ is greater than 1.

\subsection{Energy cascade rate}
\begin{figure}[hbt!]
  \centering  \includegraphics[width=1.0\linewidth]{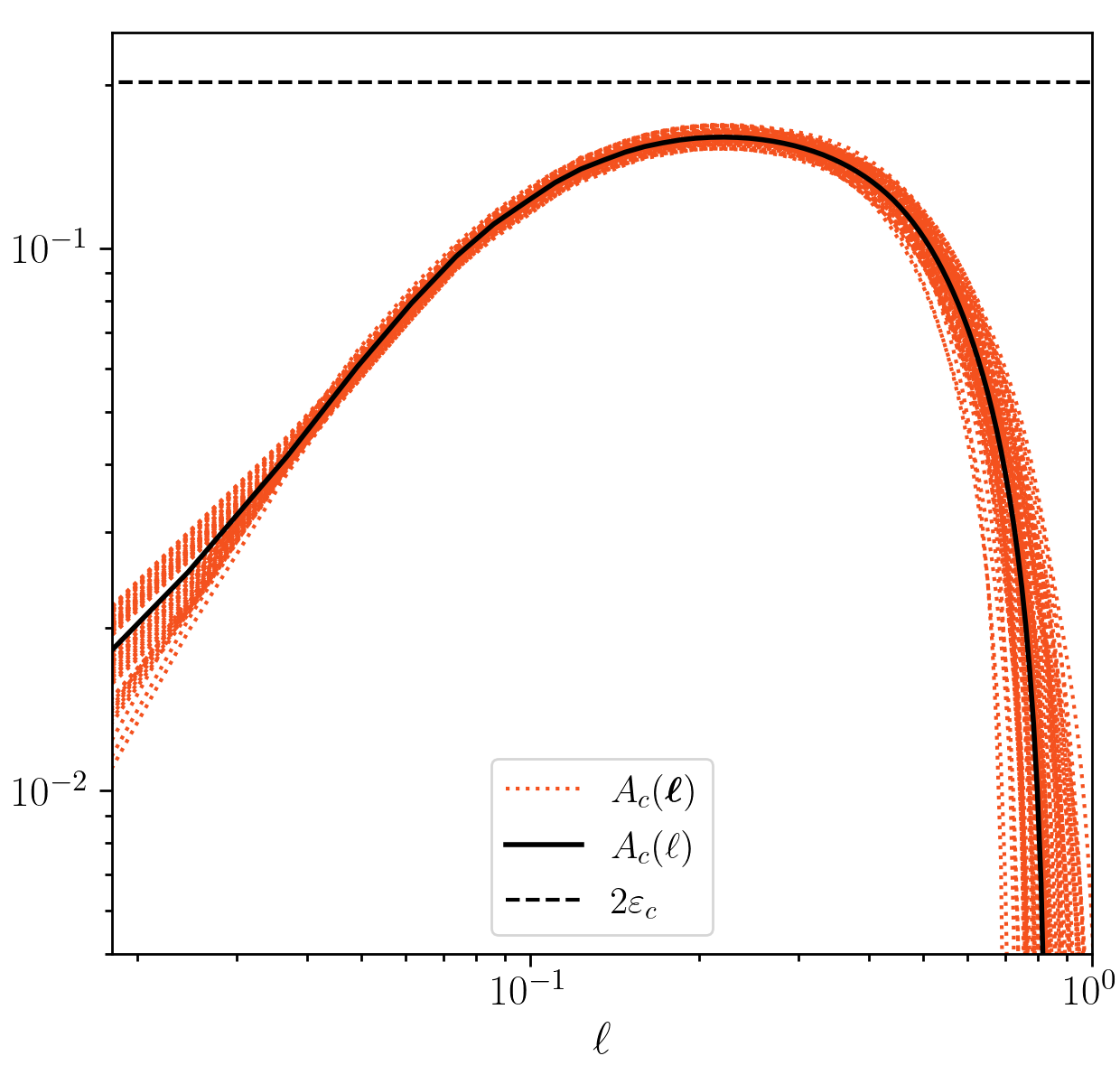}
\caption{Energy cascade rate along all the 73 directions (dotted curves) and the average over all the directions (solid black curve) for a typical simulation (Run 3a of Table \ref{tab3}).}
\label{all_direcs2}
\end{figure}

The energy cascade rate for a fully developed turbulent flow of critically balanced system $ A_c(\bm{\ell})$ is expected to show a flat region as a function of the increment vector $\bm{\ell}$ inside the inertial range. Therefore, inside the inertial range, the average injection rate or average dissipation rate is equal to the energy flux rate .i.e, $ A_c({\ell}) = const = 2\varepsilon_c = 2 D_c$. $A_c({\ell})$ is the average across all possible pairs. We have calculated the exact law with the change in the increment vector along the 73 directions, which are shown in Fig~\ref{all_direcs2}. For this , $A_c(\ell) \simeq A_c(\bm{\ell})$. Hence, $A_c(\bm{\ell})$ is shown to be approximately equal in all 73 directions.

\section{\label{sec:level5} Results and Discussion}

\begin{figure*}[hbt!]
  \centering  \includegraphics[width=0.92\linewidth]{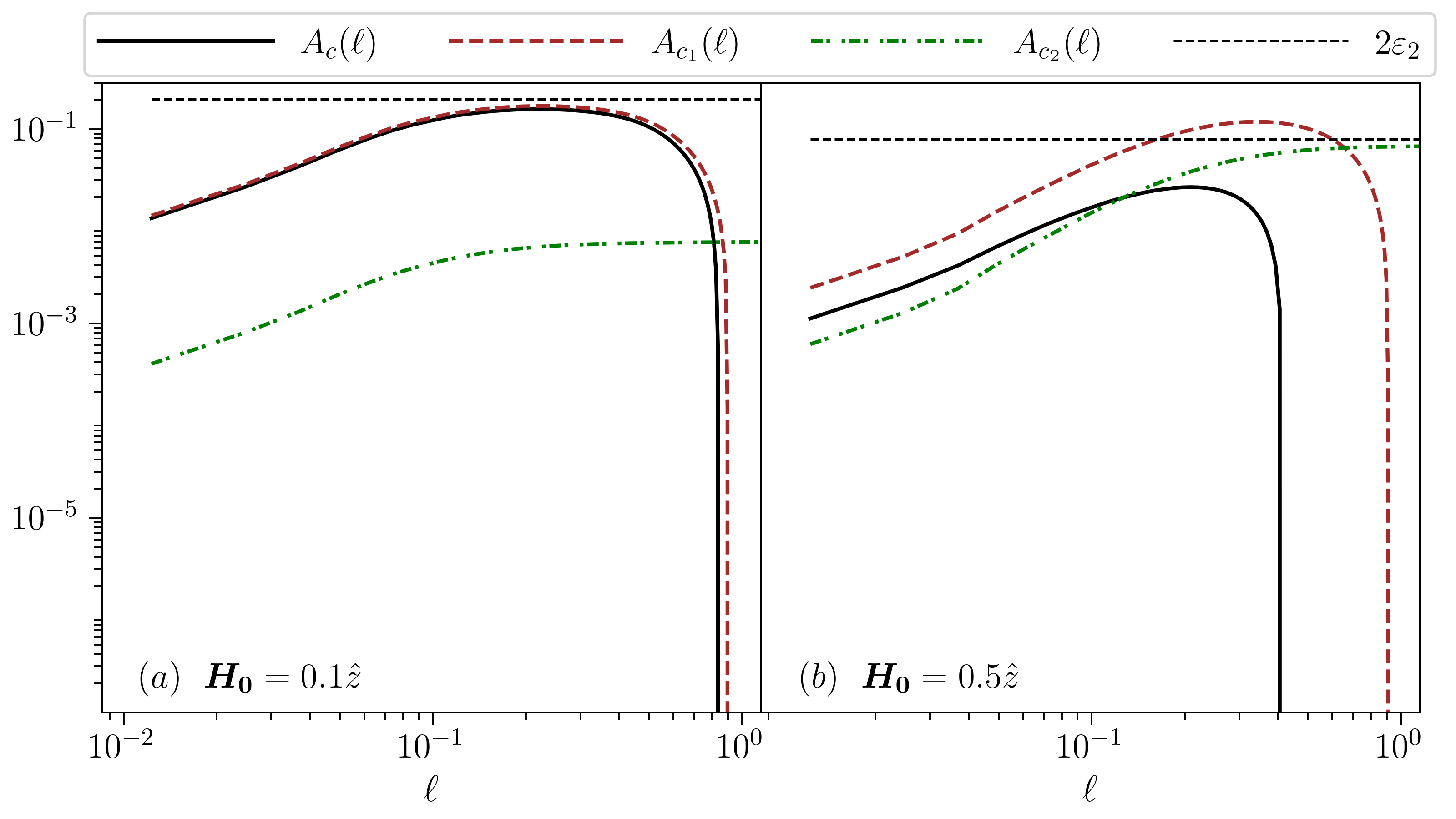}
\caption{Energy cascade rates as a function of $\ell$ for a box with $512^3$ grid points for ${\bf H_0} = 0.1 \hat{z}$ (left) and ${\bf H_0} = 0.5 \hat{z}$ (right).}
\label{flux2}
\end{figure*}

\begin{figure*}[hbt!]
  \centering  \includegraphics[width=0.92\linewidth]{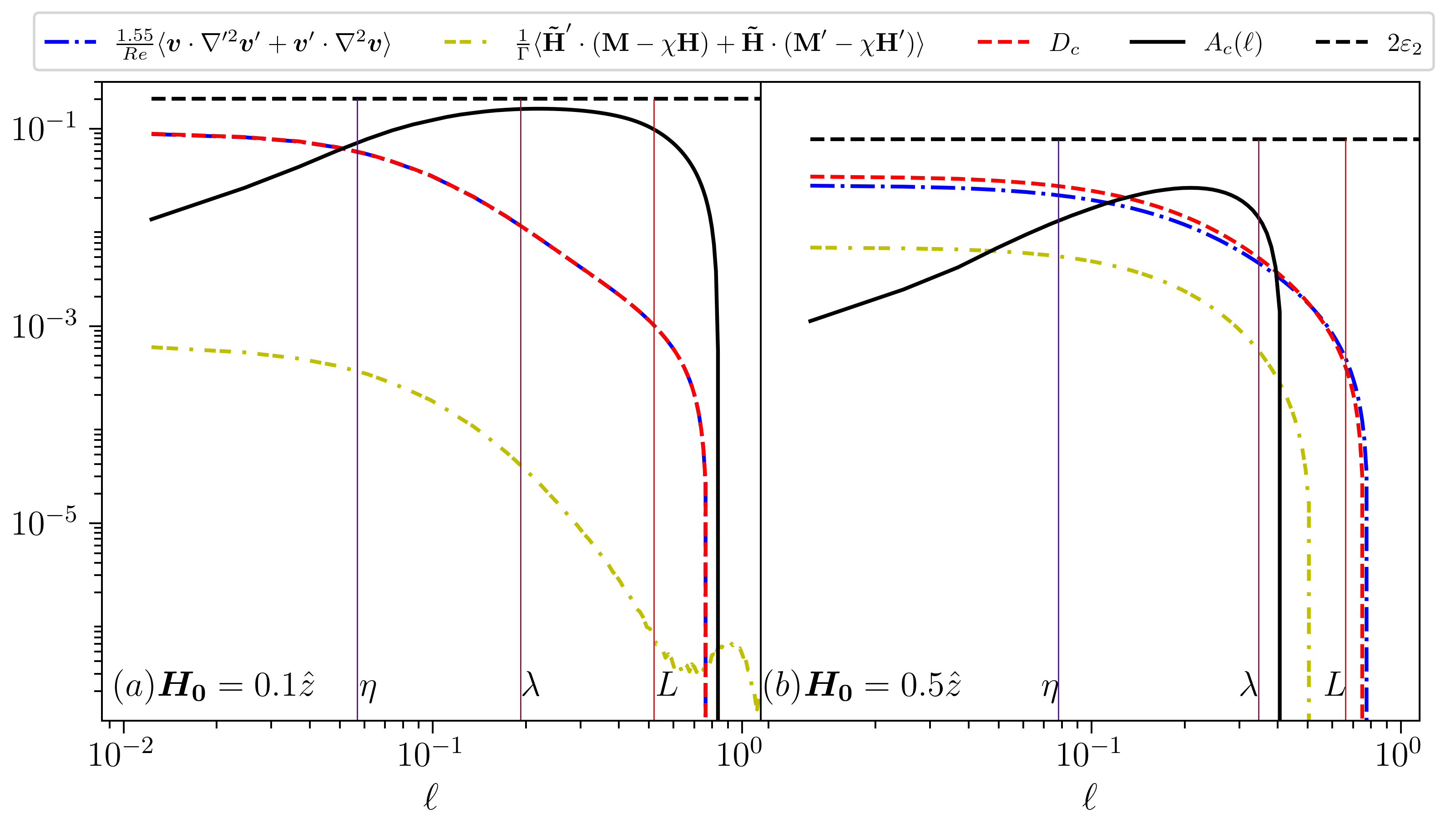}
\caption{Various components of two point energy dissipation rate and $A(\ell)$ as a function of $\ell$ for a box with $512^3$ grid points for ${\bf H_0} = 0.1 \hat{z}$ (left) and ${\bf H_0} = 0.5 \hat{z}$ (right).}
\label{diss2}
\end{figure*}

\begin{figure*}[hbt!]
  \centering
\includegraphics[width=1.0\linewidth]{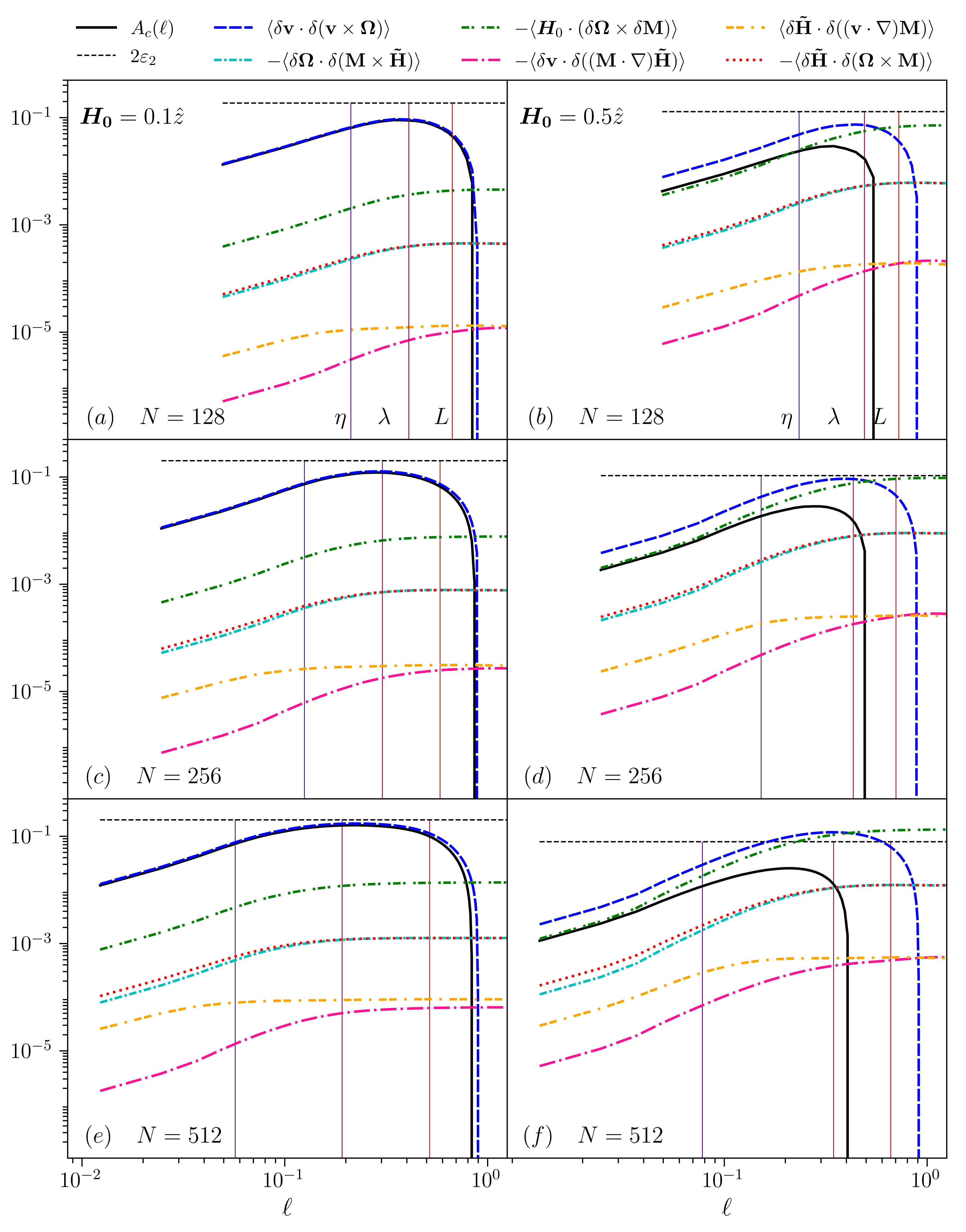}
\caption{Energy cascade rates as a function of $\ell$ for grid size $128^3$, $256^3$ and $512^3$ with an increment of the magnetic field from left to right for each grid size.}
\label{c5fig:18}
\end{figure*}

In Fig.~\ref{flux2}, we plot $A_{c_1}(\ell)$, $A_{c_2}(\ell)$, and $A(\ell)$ for the external field strengths ${H}_0 = 0.1$ and $0.5$. As can be seen in the figure, $A_c(\ell)$ is the total of all the flux contributions. Here, $A_{c_1}(\ell)$ contains the flux contribution from fluctuation of $\textbf{\~H}$, $A_{c_2}(\ell)$ includes the flux contributions owing to the external field $\textbf{H}_0$. For a fully developed turbulent flow, inside the inertial range, we aim to observe a flat region in the exact expressions $A_c(\bm{\ell})$ when plotted against the increment vector $\bm{\ell}$. The flat region indicates the scale-independent nature of the energy cascade rate, i.e., $A_c(\bm{\ell}) = const = 2\varepsilon_{c}$. It is evident from the figure that $A(\ell)$ (black solid curve) for the intermediate range has a flat region in $A(\ell)$ (black solid curve) for $H_0 = 0.1$ while for $H_0 = 0.5$ has a very sharp flat region which indicates a short-range scale independent energy cascade rate. For the critically balanced case we have a scale-independent energy cascade only for $H_0 = 0.1$. Hence, a Kolmogorov-type of universality can be found only for weak external fields where the kinetic term ($A_{c_1}(\ell)$) is only dominating in the energy cascade. For the high field the mean field term ($A_{c_2}(\ell)$) contributes to the energy cascade as a result there is a very sharp suppression of turbulence in comparison to the case where only small ferrofluid particle is taking into account. For $H_0 = 0.1$ $A_c(\ell) \approx 2 \varepsilon_c $ which holds good for Eq. \eqref{eq27} whereas for $H_0 = 0.5$ its quite different. One explanation for this might be that scale-dependent dissipative effects become non-negligible in the inertial length scales. In Fig. \ref{diss2} we plot the dissipative terms. As it is evident from the figure (a) and (b) at high field the dissipative effect is no longer negligible in the inertial range. The potential cause for this dominance of dissipative effect in the inertial range at a high field is a regular leakage of energy in terms of dissipation from a high scale to a small scale. To better understand the high field effect go to section \ref{chapter5:sec5.8}.

\begin{figure}[hbt!]
  \centering
\includegraphics[width=1.0\linewidth]{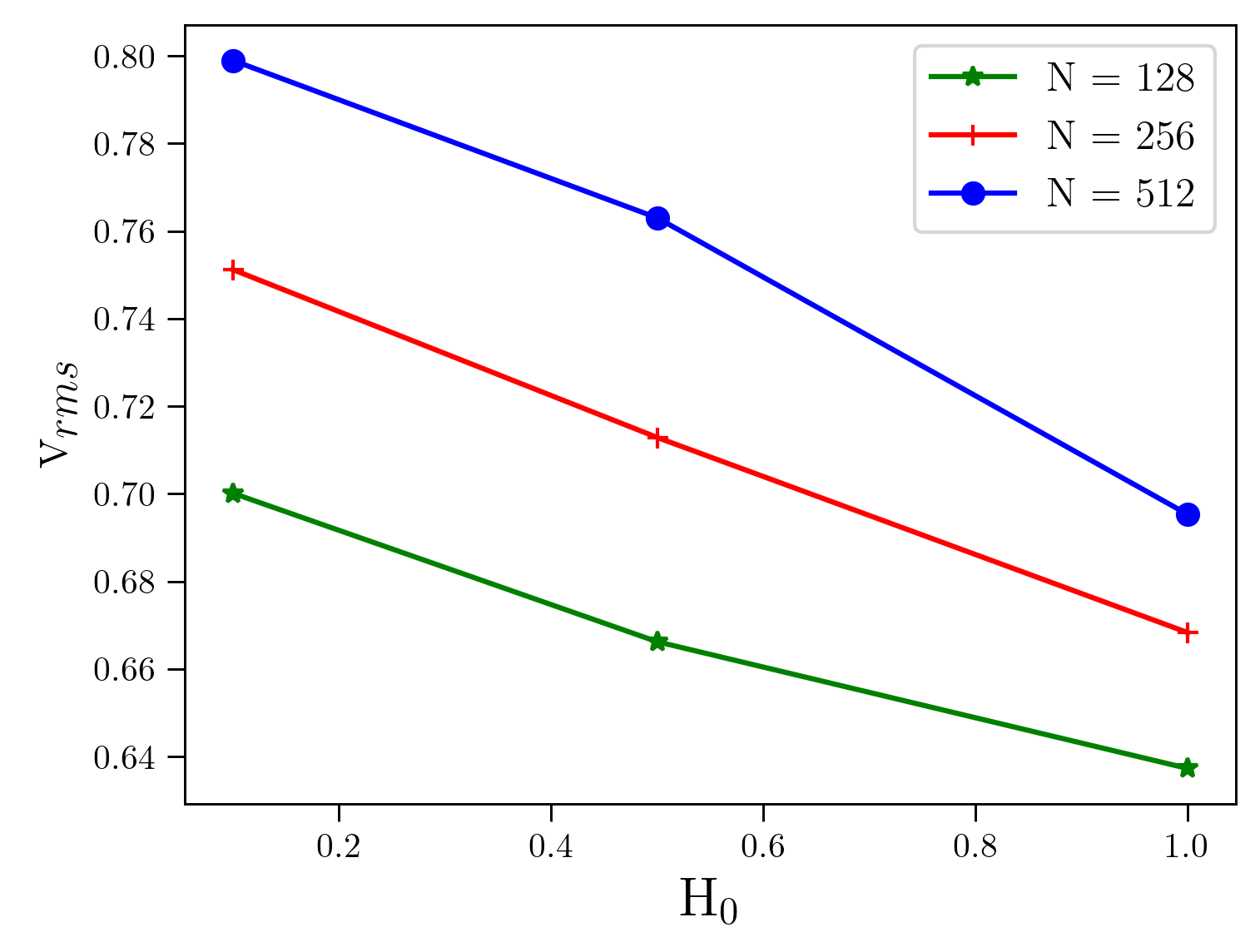}
\caption{The behaviour of the root mean square velocity with the external field.}
\label{c5fig:velo}
\end{figure}

\begin{figure}[hbt!]
  \centering
\includegraphics[width=1.0\linewidth]{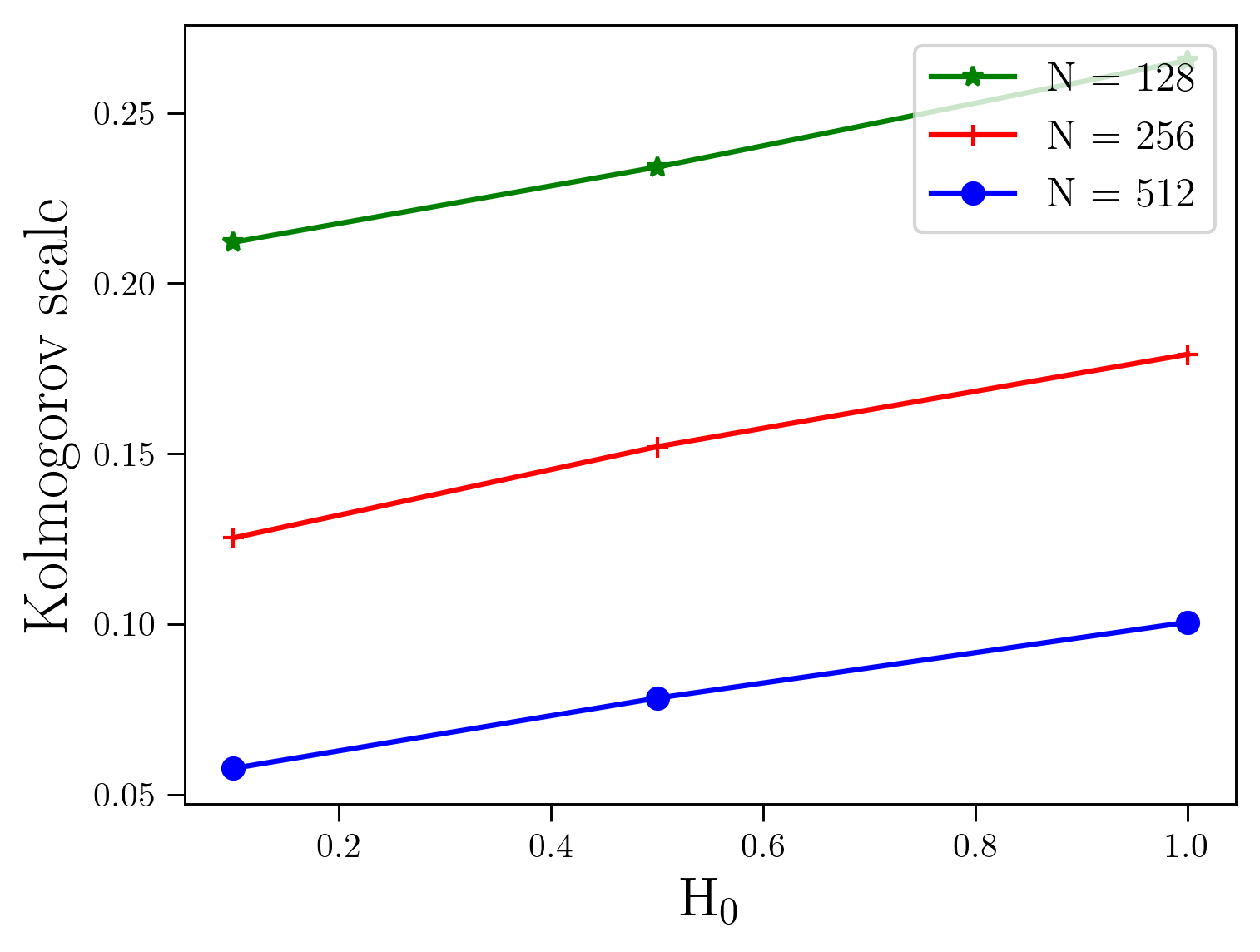}
\caption{The behaviour of Kolmogorov length with the external field.}
\label{c5fig:len1}
\end{figure}

\begin{figure}[hbt!]
  \centering
\includegraphics[width=1.0\linewidth]{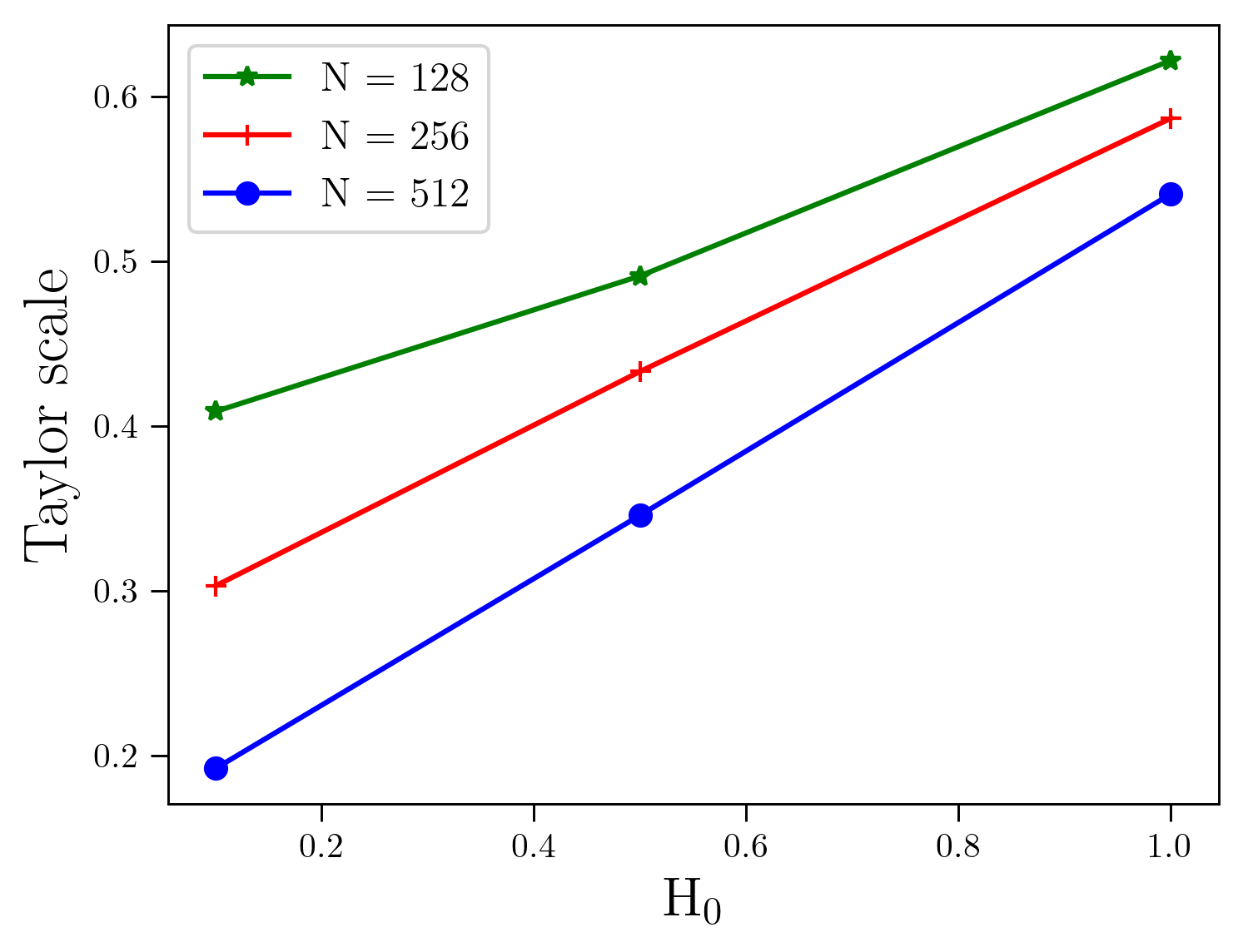}
\caption{The behaviour of Taylor length with the external field.}
\label{c5fig:len2}
\end{figure}

\begin{figure}[hbt!]
  \centering
\includegraphics[width=1.0\linewidth]{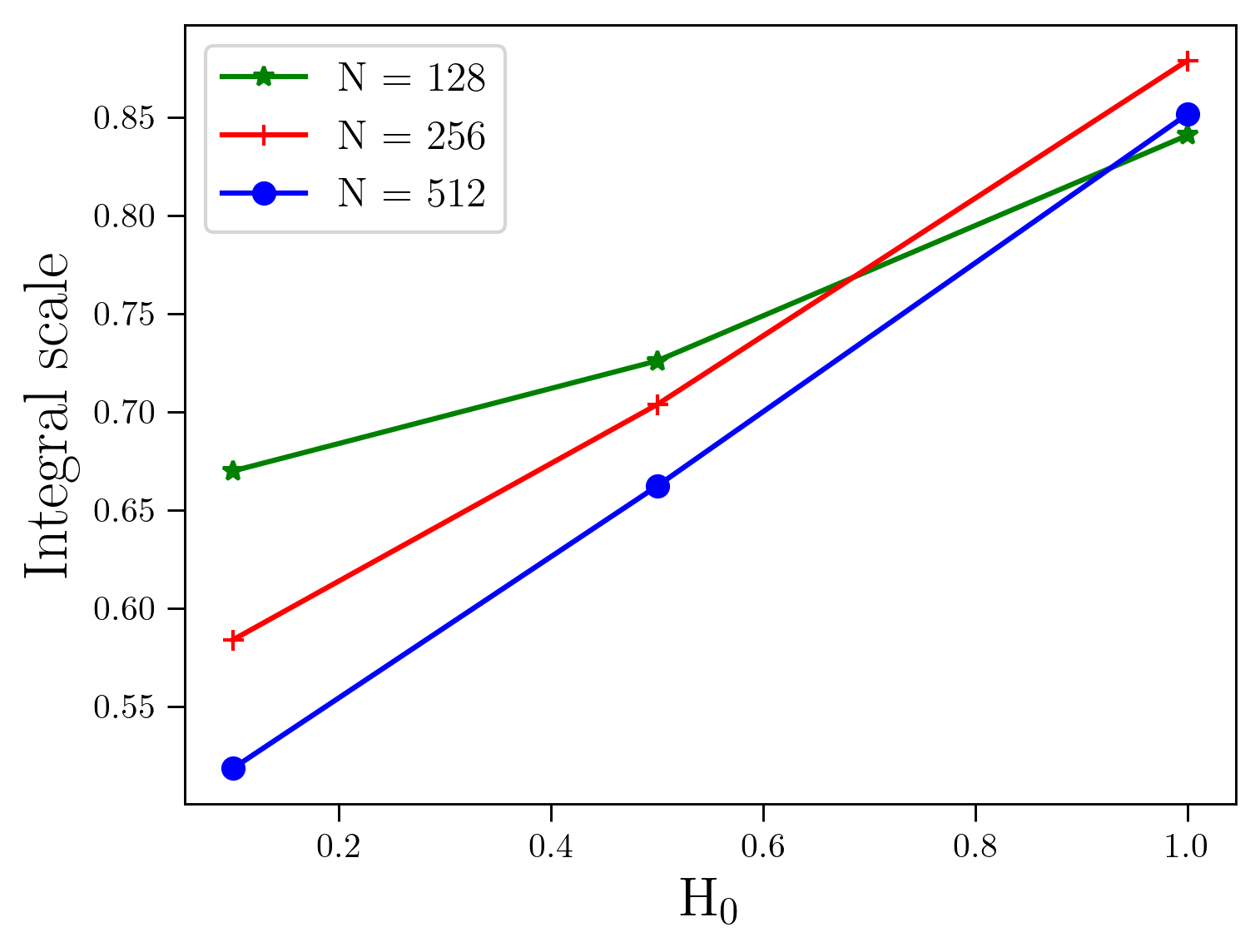}
\caption{The behaviour of integral length with the external field.}
\label{c5fig:len3}
\end{figure}


In Fig. \ref{c5fig:18} the three verticle scales indicate the largest energy-containing integral scale ($L$), inertial scale $\lambda$ and Kolmogorov scale $\eta$. It is clear from the figure that the ratio $L/\eta$ increases with the grid resolution for both external fields $H_0 = 0.1$ and $0.5$ which correspond to a wider intermediate scale. With the increase of resolution size, the gap between these length scales increases indicating the broadening of the inertial range. 
All contributing terms and the overall energy cascade rate $\varepsilon_c$ (as determined by Eq.~\eqref{29} and \eqref{30}) are plotted against the increment vector in Fig.~\ref{c5fig:18}. The average injection rate ($\varepsilon_c$) is shown as a solid black line. In Fig. \ref{c5fig:18}, part (b), (d), and (f) of each grid size is done for the Runs of Table \ref{tab3} $1b$, $2b$ and $3b$, while parts (a), (c) and (e) have been simulated by reducing mean-field respectively for each grid size. From Fig. \ref{c5fig:18}, one can observe that at the smaller value of the mean field (see parts a, c and e of each grid size), the hydrodynamic term dominates in the cascade rates. When the value of $\textbf{H}_0$ rises, various terms contribute to cascading rates. However, when we increase from a moderate value, we notice that different terms primarily contribute to the energy cascading process, and the total cascade rate is significantly outside of the dissipation range. An increase in $\textbf{H}_0$ causes the fluxes to behave differently, resulting in a transition from turbulent to laminar flow. The details about the high mean field effect are discussed in the next section \ref{chapter5:sec5.8}. The effects of the magnetic field strength on
the root-mean-square values of velocity for different grid size can be seen in Fig.~\ref{c5fig:velo}. It is clear from the figure that the fluid's velocity reduces as the external field increases, supporting the theoretical prediction that the ferromagnetic particles create a chain-like structure that obstructs the flow and causes it to decrease.
Again, the effects of the magnetic field strength on different length scales can be observed in Fig.~\ref{c5fig:len1}, Fig.~\ref{c5fig:len2} and Fig.~\ref{c5fig:len3}. It is evident from the figure that on increasing the external field the Kolmogorov, Taylor and integral length scales increases. Schumacher et. al. \cite{Schumacher2008} reported the same phenomenon, which our data confirms.

\subsection{Energy spectrum}

\begin{figure*}[hbt!]
     \centering  \includegraphics[width=0.92\linewidth]{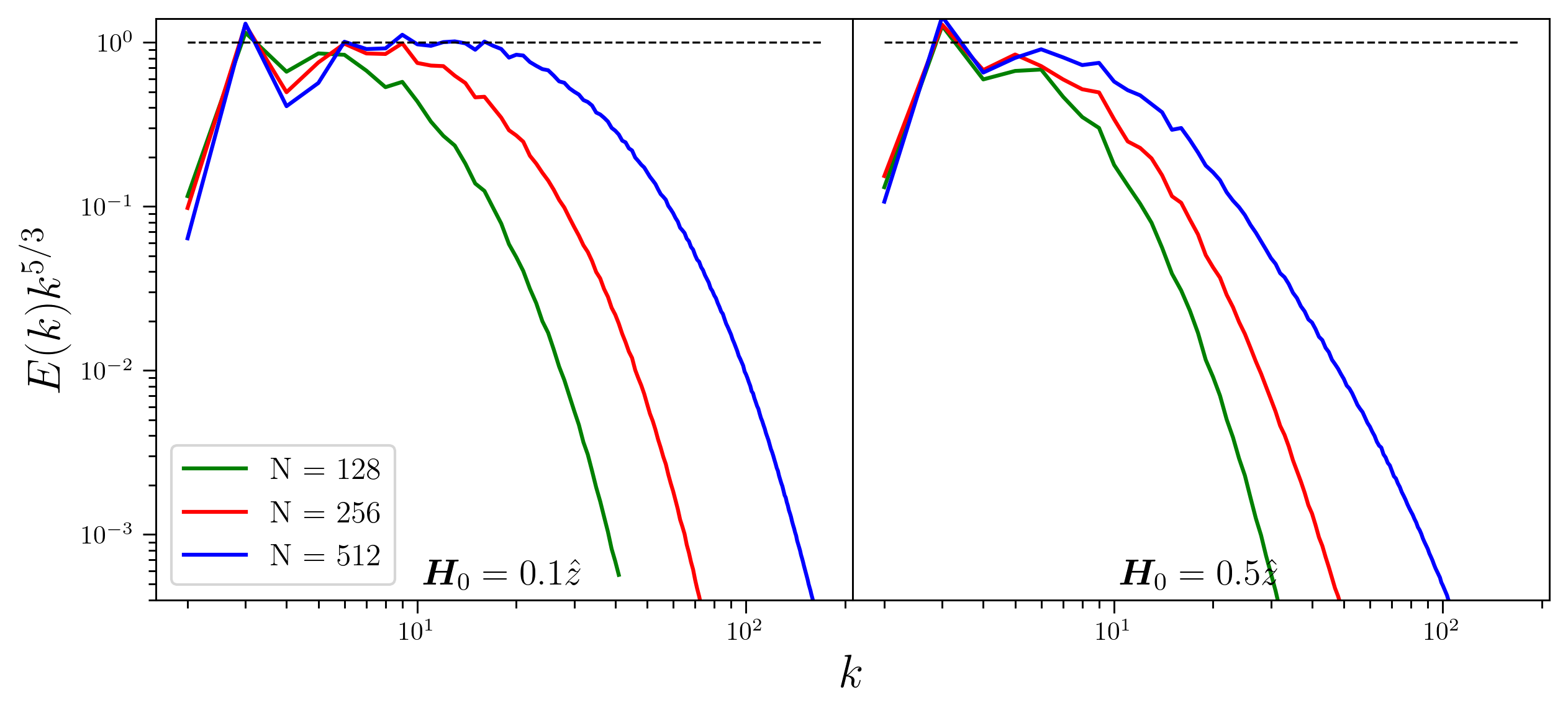}
         \caption{Compensated kinetic energy spectrum for all Runs for ${H}_0 = 0.1$ and ${H}_0=0.5$ in Table \ref{tab3}.}
         \label{c5fig:Spectrum2}
\end{figure*}

Fig.~\ref{c5fig:Spectrum2} shows the kinetic energy spectrum. According to the figure, we obtain a $k^{-5/3}$ spectrum for $H_0 = 0.1$, which indicates ferrofluid turbulence with an energy cascade associated with the Kolmogorov type. The figure shows that the spectrum width decreases as $H_0$ increases. It is evident from the figure that at $H_0 = 0.5$ there is a very small range of $k^{-5/3}$ spectrum which indicates the dissipative effect in the inertial range as explained in the earlier section. Again, in this case, it is impossible to establish a spectrum for total energy since the exact differential form of work caused by magnetization cannot be expressed.

\subsection{\label{chapter5:sec5.8}Effect of strong background magnetic field}
\begin{figure*}[hbt!]
         \centering                  
         \includegraphics[width=0.9\linewidth]{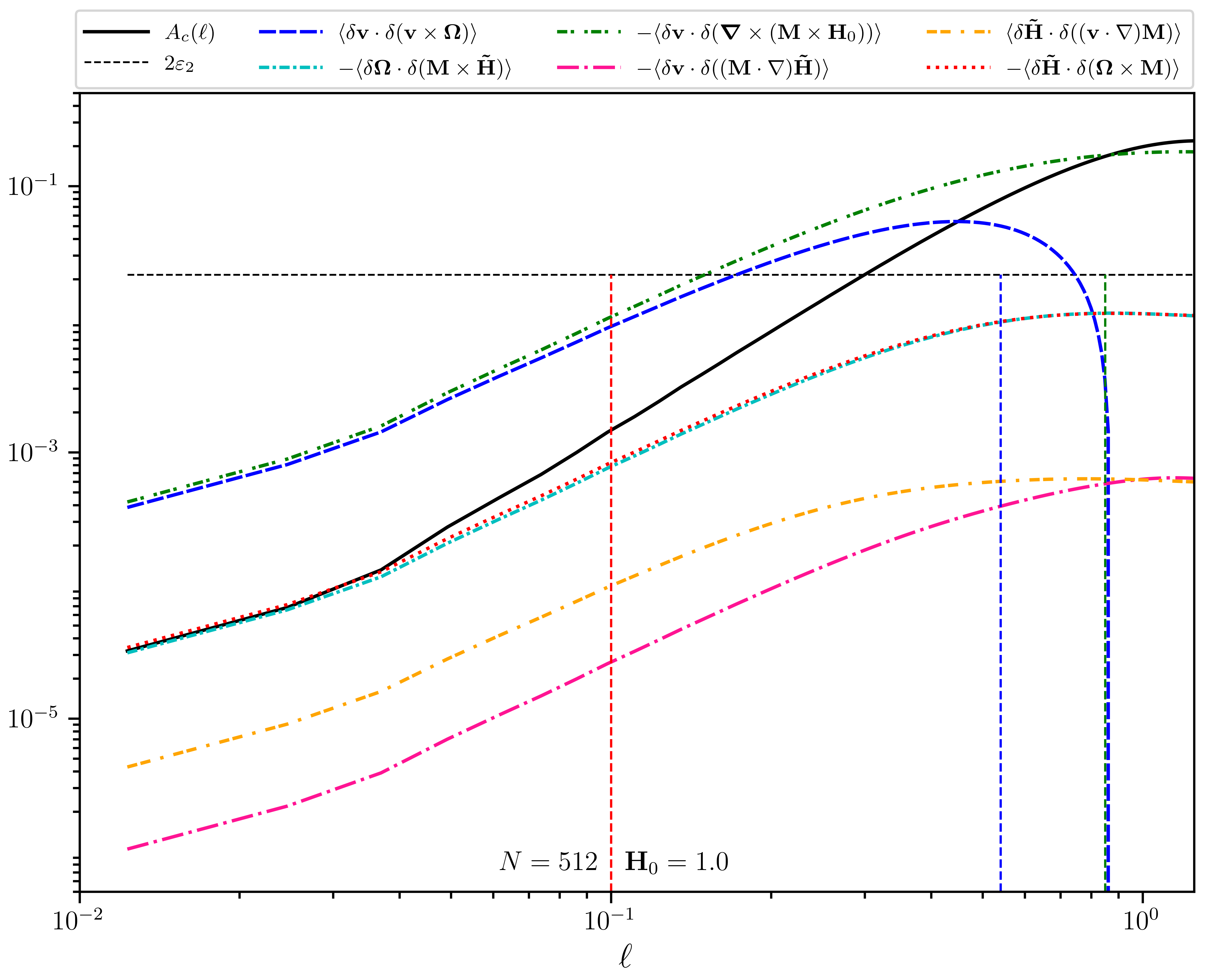}
         \label{fig:8b}
\caption{Energy cascade rate as a function of $\ell$ at $t = 10$ for $512^3$.}
\label{c7fig:19}
\end{figure*}

\begin{figure}[hbt!]
     \centering  \includegraphics[width=1.0\linewidth]{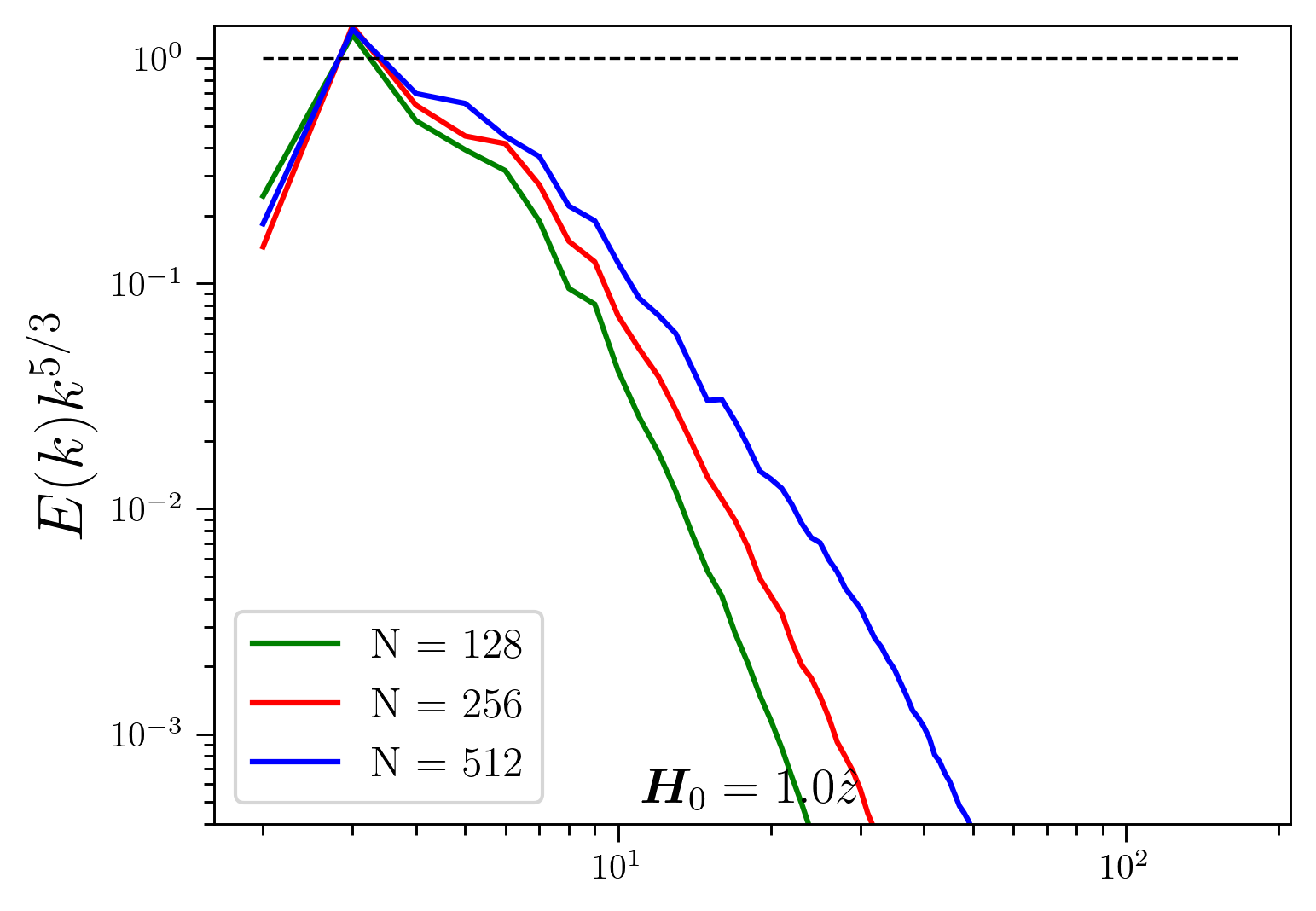}
         \caption{Compensated kinetic energy spectrum for grid size of $\textbf{H}_0 = 1.0 $ in Table \ref{tab3}.}
         \label{c5fig:Spectrum3}
\end{figure}

In the absence of an external magnetic field, the magnetic particles of ferrofluid are free to move and show Brownian motion. According to the theory when we apply a mean magnetic field the magnetic moment of the magnetic particle tries to align towards the direction of the magnetic field and form a chain-like structure. As we increase the mean field the chain-like structure grows and restricts the particles from flowing \citep{Valentin2004}. Therefore, at a high mean field, the flow of fluid transits to a laminar-type structure. To get a turbulence at high mean field the velocity should be very high accordingly. In Fig.~\ref{c5fig:15} we can see the snapshot of velocity at the high mean magnetic field. In the velocity snapshot, the eddies are disappearing and can see the patches which prefer the laminar-type flow. Further, in the energy cascade rate (Fig.~\ref{c7fig:19}), the slope of fluxes decreases continuously with the decrease in length scale (large scale to small scale). There is no flat region in the inertial range. Hence. the flow of ferrofluid transit from turbulent to laminar flow on the increase of mean-field. This effect can also be seen in the energy spectrum (see Fig.~\ref{c5fig:Spectrum3}) where a continuous negative slope can be observed with the increase of wavenumber. It is impossible to see the flat region or inertial region where a constant cascade of energy is occurring. Hence, it is clear from Fig.~\ref{c7fig:19} and \ref{c5fig:Spectrum3} that at high-field, we do not obtain the usual Kolmogorov turbulence for the critically balanced case. 

\textbf{Energy spectrum}:
The kinetic energy spectrum also verifies that at a high external field, the continuous energy cascade cannot be observed (see Fig. \ref{c5fig:Spectrum3}). It can be clearly observed from the figure that there is a continuous cascade of energy through all length scales. Hence, one cannot identify an inertial range of scale.

\section{\label{sec:level6}Conclusions and Summary}

Using two-point statistics, we have calculated the energy flux rate in homogeneous turbulence for ferrofluids in critically balanced flow. We have investigated the statistical dynamics of the different terms that are contributing to the energy cascade rate. The terms $\delta (\textbf{v} \times \boldsymbol{\Omega}) \cdot \delta \textbf{v}$ and $\textbf{H}_0\cdot (\delta \boldsymbol{\Omega} \times \delta \textbf{M})$ mostly contributes to the energy cascade from inertial to Kolmogorov range while the contribution of other terms is negligibly small (see Fig. \ref{c5fig:18}).

The numerical outcomes indicate that with an increase in spatial resolution, the overall energy flux rate  $\varepsilon_c$ tends to approach the dissipation rate. Synchronization between particle rotation and flow vorticity have reduced shear stress and energy dissipation at certain scales. This dampens turbulence locally, depending on the strength of the external magnetic field and flow conditions. Further, the magnetic particles no longer experience a torque due to the viscous coupling with the fluid. Once again, it is evident from the analysis of Fig. \ref{c5fig:18} that a reduction in the intensity of the external field leads to the emergence of kinetic energy as a prominent contributor to the total energy cascade. Simultaneously, the impact of other components diminishes, resulting in a gradual relaxing process. Furthermore, when the strength of the magnetic field increases, the magnetization relaxation becomes dominant over the kinetic component, leading to a decrease in the turbulent behaviour. The application of strong magnetic field gradients offers a means to suppress turbulence within ferrofluids.

A Kolmogorov-type cascade for kinetic energy in ferrofluid turbulence is justified by the kinetic energy power spectrum, 
which provides a $k^{-5/3}$ spectrum for $H_0 = 0.1$ and $0.5$. However, at high field $H_0 = 1.0$, this Kolmogorov-type phenomenon cannot be observed which indicates that turbulence cannot be developed due to the formation of chain-like structure. Therefore, a significant amount of injection energy is needed to maintain a turbulent flow.

We can study the evolution of thermal effect when the particle rotation rate becomes equal to the vorticity of the flow which may affect transport properties such as heat and mass transfer \cite{BAFinlayson1970, Mouraya2023}. Additionally, the theory of vanishing nonlinear transfer can be employed for the critically balanced case to investigate turbulence relaxation \cite{Banerjee2023PVNLT, pan2024universal}. Using this approach, a study on the energy cascade process in the spectrum space in terms of the triad interactions may verify the universality by determining the associated spectral flux rates and investigating the shell-to-shell transfer rates \cite{Banerjee2023fundamental}.

\section*{Acknowledgments}
The simulation code is developed with the help of parallelization schemes given in Ref. \citep{Mortensen2016}. The simulations are performed using the support and resources provided by PARAM Sanganak under the National Supercomputing Mission, Government of India at the Indian Institute of Technology, Kanpur.

\pagebreak
	
\bibliography{main}

\providecommand{\noopsort}[1]{}\providecommand{\singleletter}[1]{#1}%
\begin{thebibliography}{56}%
\makeatletter
\providecommand \@ifxundefined [1]{%
 \@ifx{#1\undefined}
}%
\providecommand \@ifnum [1]{%
 \ifnum #1\expandafter \@firstoftwo
 \else \expandafter \@secondoftwo
 \fi
}%
\providecommand \@ifx [1]{%
 \ifx #1\expandafter \@firstoftwo
 \else \expandafter \@secondoftwo
 \fi
}%
\providecommand \natexlab [1]{#1}%
\providecommand \enquote  [1]{``#1''}%
\providecommand \bibnamefont  [1]{#1}%
\providecommand \bibfnamefont [1]{#1}%
\providecommand \citenamefont [1]{#1}%
\providecommand \href@noop [0]{\@secondoftwo}%
\providecommand \href [0]{\begingroup \@sanitize@url \@href}%
\providecommand \@href[1]{\@@startlink{#1}\@@href}%
\providecommand \@@href[1]{\endgroup#1\@@endlink}%
\providecommand \@sanitize@url [0]{\catcode `\\12\catcode `\$12\catcode `\&12\catcode `\#12\catcode `\^12\catcode `\_12\catcode `\%12\relax}%
\providecommand \@@startlink[1]{}%
\providecommand \@@endlink[0]{}%
\providecommand \url  [0]{\begingroup\@sanitize@url \@url }%
\providecommand \@url [1]{\endgroup\@href {#1}{\urlprefix }}%
\providecommand \urlprefix  [0]{URL }%
\providecommand \Eprint [0]{\href }%
\providecommand \doibase [0]{https://doi.org/}%
\providecommand \selectlanguage [0]{\@gobble}%
\providecommand \bibinfo  [0]{\@secondoftwo}%
\providecommand \bibfield  [0]{\@secondoftwo}%
\providecommand \translation [1]{[#1]}%
\providecommand \BibitemOpen [0]{}%
\providecommand \bibitemStop [0]{}%
\providecommand \bibitemNoStop [0]{.\EOS\space}%
\providecommand \EOS [0]{\spacefactor3000\relax}%
\providecommand \BibitemShut  [1]{\csname bibitem#1\endcsname}%
\let\auto@bib@innerbib\@empty
\bibitem [{\citenamefont {Finlayson}(1970)}]{BAFinlayson1970}%
  \BibitemOpen
  \bibfield  {author} {\bibinfo {author} {\bibfnamefont {B.~A.}\ \bibnamefont {Finlayson}},\ }\bibfield  {title} {\bibinfo {title} {Convective instability of ferromagnetic fluids},\ }\href@noop {} {\bibfield  {journal} {\bibinfo  {journal} {Journal of Fluid Mechanics}\ }\textbf {\bibinfo {volume} {40}},\ \bibinfo {pages} {753} (\bibinfo {year} {1970})}\BibitemShut {NoStop}%
\bibitem [{\citenamefont {Rosensweig}(1997)}]{Rosensweig1997}%
  \BibitemOpen
  \bibfield  {author} {\bibinfo {author} {\bibfnamefont {R.}~\bibnamefont {Rosensweig}},\ }\href@noop {} {\emph {\bibinfo {title} {Ferrohydrodynamics}}}\ (\bibinfo  {publisher} {Cambridge University Press},\ \bibinfo {address} {Dover Publications, inc. Mineola, New York},\ \bibinfo {year} {1997})\BibitemShut {NoStop}%
\bibitem [{\citenamefont {Schumacher}\ \emph {et~al.}(2008)\citenamefont {Schumacher}, \citenamefont {Riley},\ and\ \citenamefont {Finlayson}}]{Schumacher2008}%
  \BibitemOpen
  \bibfield  {author} {\bibinfo {author} {\bibfnamefont {K.~R.}\ \bibnamefont {Schumacher}}, \bibinfo {author} {\bibfnamefont {J.~J.}\ \bibnamefont {Riley}},\ and\ \bibinfo {author} {\bibfnamefont {B.~A.}\ \bibnamefont {Finlayson}},\ }\bibfield  {title} {\bibinfo {title} {Homogeneous turbulence in ferrofluids with a steady magnetic field},\ }\href@noop {} {\bibfield  {journal} {\bibinfo  {journal} {J. Fluid Mech.}\ }\textbf {\bibinfo {volume} {599}},\ \bibinfo {pages} {1} (\bibinfo {year} {2008})}\BibitemShut {NoStop}%
\bibitem [{\citenamefont {Shokrollahi}(2013)}]{Shokrollahi2013}%
  \BibitemOpen
  \bibfield  {author} {\bibinfo {author} {\bibfnamefont {H.}~\bibnamefont {Shokrollahi}},\ }\bibfield  {title} {\bibinfo {title} {Structure, synthetic methods, magnetic properties and biomedical applications of ferrofluids},\ }\href {https://doi.org/https://doi.org/10.1016/j.msec.2013.03.028} {\bibfield  {journal} {\bibinfo  {journal} {Materials Science and Engineering: C}\ }\textbf {\bibinfo {volume} {33}},\ \bibinfo {pages} {2476} (\bibinfo {year} {2013})}\BibitemShut {NoStop}%
\bibitem [{\citenamefont {Hang~Koh}\ \emph {et~al.}(2013)\citenamefont {Hang~Koh}, \citenamefont {Seng~Lok},\ and\ \citenamefont {Nguyen}}]{Koh2013}%
  \BibitemOpen
  \bibfield  {author} {\bibinfo {author} {\bibfnamefont {W.}~\bibnamefont {Hang~Koh}}, \bibinfo {author} {\bibfnamefont {K.}~\bibnamefont {Seng~Lok}},\ and\ \bibinfo {author} {\bibfnamefont {N.-T.}\ \bibnamefont {Nguyen}},\ }\bibfield  {title} {\bibinfo {title} {{A Digital Micro Magnetofluidic Platform For Lab-on-a-Chip Applications}},\ }\href {https://doi.org/10.1115/1.4023443} {\bibfield  {journal} {\bibinfo  {journal} {Journal of Fluids Engineering}\ }\textbf {\bibinfo {volume} {135}},\ \bibinfo {pages} {021302} (\bibinfo {year} {2013})}\BibitemShut {NoStop}%
\bibitem [{\citenamefont {Jing}\ \emph {et~al.}(2020)\citenamefont {Jing}, \citenamefont {Sun}, \citenamefont {Jin}, \citenamefont {Thangamuthu},\ and\ \citenamefont {Tang}}]{Jing2021}%
  \BibitemOpen
  \bibfield  {author} {\bibinfo {author} {\bibfnamefont {D.}~\bibnamefont {Jing}}, \bibinfo {author} {\bibfnamefont {L.}~\bibnamefont {Sun}}, \bibinfo {author} {\bibfnamefont {J.}~\bibnamefont {Jin}}, \bibinfo {author} {\bibfnamefont {M.}~\bibnamefont {Thangamuthu}},\ and\ \bibinfo {author} {\bibfnamefont {J.}~\bibnamefont {Tang}},\ }\bibfield  {title} {\bibinfo {title} {Magneto-optical transmission in magnetic nanoparticle suspensions for different optical applications: a review},\ }\href {https://doi.org/10.1088/1361-6463/abb8fd} {\bibfield  {journal} {\bibinfo  {journal} {Journal of Physics D: Applied Physics}\ }\textbf {\bibinfo {volume} {54}},\ \bibinfo {pages} {013001} (\bibinfo {year} {2020})}\BibitemShut {NoStop}%
\bibitem [{\citenamefont {Chung}\ \emph {et~al.}(2021)\citenamefont {Chung}, \citenamefont {Parsons},\ and\ \citenamefont {Zheng}}]{Chung2021}%
  \BibitemOpen
  \bibfield  {author} {\bibinfo {author} {\bibfnamefont {H.-J.}\ \bibnamefont {Chung}}, \bibinfo {author} {\bibfnamefont {A.~M.}\ \bibnamefont {Parsons}},\ and\ \bibinfo {author} {\bibfnamefont {L.}~\bibnamefont {Zheng}},\ }\bibfield  {title} {\bibinfo {title} {Magnetically controlled soft robotics utilizing elastomers and gels in actuation: A review},\ }\href {https://doi.org/https://doi.org/10.1002/aisy.202000186} {\bibfield  {journal} {\bibinfo  {journal} {Advanced Intelligent Systems}\ }\textbf {\bibinfo {volume} {3}},\ \bibinfo {pages} {2000186} (\bibinfo {year} {2021})}\BibitemShut {NoStop}%
\bibitem [{\citenamefont {Shliomis}(1972)}]{Shliomis1972}%
  \BibitemOpen
  \bibfield  {author} {\bibinfo {author} {\bibfnamefont {M.~I.}\ \bibnamefont {Shliomis}},\ }\bibfield  {title} {\bibinfo {title} {Effective viscosity of magnetic suspensions},\ }\href@noop {} {\bibfield  {journal} {\bibinfo  {journal} {Sov. Phys. JETP}\ }\textbf {\bibinfo {volume} {34}},\ \bibinfo {pages} {1291} (\bibinfo {year} {1972})}\BibitemShut {NoStop}%
\bibitem [{\citenamefont {Zebib}(1996)}]{Zebib1996}%
  \BibitemOpen
  \bibfield  {author} {\bibinfo {author} {\bibfnamefont {A.}~\bibnamefont {Zebib}},\ }\bibfield  {title} {\bibinfo {title} {Thermal convection in a magnetic fluid},\ }\href {https://doi.org/10.1017/S0022112096007665} {\bibfield  {journal} {\bibinfo  {journal} {Journal of Fluid Mechanics}\ }\textbf {\bibinfo {volume} {321}},\ \bibinfo {pages} {121} (\bibinfo {year} {1996})}\BibitemShut {NoStop}%
\bibitem [{\citenamefont {Kaloni}\ and\ \citenamefont {Lou}(2004)}]{Kaloni2004}%
  \BibitemOpen
  \bibfield  {author} {\bibinfo {author} {\bibfnamefont {P.~N.}\ \bibnamefont {Kaloni}}\ and\ \bibinfo {author} {\bibfnamefont {J.~X.}\ \bibnamefont {Lou}},\ }\bibfield  {title} {\bibinfo {title} {Convective instability of magnetic fluids},\ }\href {https://doi.org/10.1103/PhysRevE.70.026313} {\bibfield  {journal} {\bibinfo  {journal} {Phys. Rev. E}\ }\textbf {\bibinfo {volume} {70}},\ \bibinfo {pages} {026313} (\bibinfo {year} {2004})}\BibitemShut {NoStop}%
\bibitem [{\citenamefont {Shliomis}\ and\ \citenamefont {Morozov}(1994)}]{Shliomis1994}%
  \BibitemOpen
  \bibfield  {author} {\bibinfo {author} {\bibfnamefont {M.~I.}\ \bibnamefont {Shliomis}}\ and\ \bibinfo {author} {\bibfnamefont {K.~I.}\ \bibnamefont {Morozov}},\ }\bibfield  {title} {\bibinfo {title} {Negative viscosity of ferrofluid under alternating magnetic field},\ }\href {https://doi.org/10.1063/1.868108} {\bibfield  {journal} {\bibinfo  {journal} {Physics of Fluids}\ }\textbf {\bibinfo {volume} {6}},\ \bibinfo {pages} {2855} (\bibinfo {year} {1994})}\BibitemShut {NoStop}%
\bibitem [{\citenamefont {Bacri}\ \emph {et~al.}(1995)\citenamefont {Bacri}, \citenamefont {Perzynski}, \citenamefont {Shliomis},\ and\ \citenamefont {Burde}}]{Bacri1995}%
  \BibitemOpen
  \bibfield  {author} {\bibinfo {author} {\bibfnamefont {J.-C.}\ \bibnamefont {Bacri}}, \bibinfo {author} {\bibfnamefont {R.}~\bibnamefont {Perzynski}}, \bibinfo {author} {\bibfnamefont {M.~I.}\ \bibnamefont {Shliomis}},\ and\ \bibinfo {author} {\bibfnamefont {G.~I.}\ \bibnamefont {Burde}},\ }\bibfield  {title} {\bibinfo {title} {``negative-viscosity'' effect in a magnetic fluid},\ }\href {https://doi.org/10.1103/PhysRevLett.75.2128} {\bibfield  {journal} {\bibinfo  {journal} {Phys. Rev. Lett.}\ }\textbf {\bibinfo {volume} {75}},\ \bibinfo {pages} {2128} (\bibinfo {year} {1995})}\BibitemShut {NoStop}%
\bibitem [{\citenamefont {Finlayson}(2013)}]{BAFinlayson2013}%
  \BibitemOpen
  \bibfield  {author} {\bibinfo {author} {\bibfnamefont {B.~A.}\ \bibnamefont {Finlayson}},\ }\bibfield  {title} {\bibinfo {title} {Spin-up of ferrofluids: The impact of the spin viscosity and the langevin function},\ }\href {https://doi.org/10.1063/1.4812295} {\bibfield  {journal} {\bibinfo  {journal} {Physics of Fluids}\ }\textbf {\bibinfo {volume} {25}},\ \bibinfo {pages} {073101} (\bibinfo {year} {2013})}\BibitemShut {NoStop}%
\bibitem [{\citenamefont {Mendelev}\ and\ \citenamefont {Ivanov}(2004)}]{Valentin2004}%
  \BibitemOpen
  \bibfield  {author} {\bibinfo {author} {\bibfnamefont {V.~S.}\ \bibnamefont {Mendelev}}\ and\ \bibinfo {author} {\bibfnamefont {A.~O.}\ \bibnamefont {Ivanov}},\ }\bibfield  {title} {\bibinfo {title} {Ferrofluid aggregation in chains under the influence of a magnetic field},\ }\href {https://doi.org/10.1103/PhysRevE.70.051502} {\bibfield  {journal} {\bibinfo  {journal} {Phys. Rev. E}\ }\textbf {\bibinfo {volume} {70}},\ \bibinfo {pages} {051502} (\bibinfo {year} {2004})}\BibitemShut {NoStop}%
\bibitem [{\citenamefont {Dixit}\ and\ \citenamefont {Pattamatta}(2020)}]{Dixit2020}%
  \BibitemOpen
  \bibfield  {author} {\bibinfo {author} {\bibfnamefont {D.~D.}\ \bibnamefont {Dixit}}\ and\ \bibinfo {author} {\bibfnamefont {A.}~\bibnamefont {Pattamatta}},\ }\bibfield  {title} {\bibinfo {title} {Effect of uniform external magnetic-field on natural convection heat transfer in a cubical cavity filled with magnetic nano-dispersion},\ }\href {https://doi.org/https://doi.org/10.1016/j.ijheatmasstransfer.2019.118828} {\bibfield  {journal} {\bibinfo  {journal} {International Journal of Heat and Mass Transfer}\ }\textbf {\bibinfo {volume} {146}},\ \bibinfo {pages} {118828} (\bibinfo {year} {2020})}\BibitemShut {NoStop}%
\bibitem [{\citenamefont {Li}\ \emph {et~al.}(2023)\citenamefont {Li}, \citenamefont {Li}, \citenamefont {Han}, \citenamefont {Li}, \citenamefont {Yan}, \citenamefont {Zhao},\ and\ \citenamefont {Chen}}]{Li2023}%
  \BibitemOpen
  \bibfield  {author} {\bibinfo {author} {\bibfnamefont {W.}~\bibnamefont {Li}}, \bibinfo {author} {\bibfnamefont {Z.}~\bibnamefont {Li}}, \bibinfo {author} {\bibfnamefont {W.}~\bibnamefont {Han}}, \bibinfo {author} {\bibfnamefont {Y.}~\bibnamefont {Li}}, \bibinfo {author} {\bibfnamefont {S.}~\bibnamefont {Yan}}, \bibinfo {author} {\bibfnamefont {Q.}~\bibnamefont {Zhao}},\ and\ \bibinfo {author} {\bibfnamefont {F.}~\bibnamefont {Chen}},\ }\bibfield  {title} {\bibinfo {title} {{Measured viscosity characteristics of Fe3O4 ferrofluid in magnetic and thermal fields}},\ }\href {https://doi.org/10.1063/5.0131551} {\bibfield  {journal} {\bibinfo  {journal} {Physics of Fluids}\ }\textbf {\bibinfo {volume} {35}},\ \bibinfo {pages} {012002} (\bibinfo {year} {2023})},\ \Eprint {https://arxiv.org/abs/https://pubs.aip.org/aip/pof/article-pdf/doi/10.1063/5.0131551/16652243/012002\_1\_online.pdf} {https://pubs.aip.org/aip/pof/article-pdf/doi/10.1063/5.0131551/16652243/012002\_1\_online.pdf} \BibitemShut {NoStop}%
\bibitem [{\citenamefont {Mouraya}\ \emph {et~al.}(2024)\citenamefont {Mouraya}, \citenamefont {Pan},\ and\ \citenamefont {Banerjee}}]{Mouraya2024}%
  \BibitemOpen
  \bibfield  {author} {\bibinfo {author} {\bibfnamefont {S.}~\bibnamefont {Mouraya}}, \bibinfo {author} {\bibfnamefont {N.}~\bibnamefont {Pan}},\ and\ \bibinfo {author} {\bibfnamefont {S.}~\bibnamefont {Banerjee}},\ }\bibfield  {title} {\bibinfo {title} {Stationary and nonstationary energy cascades in homogeneous ferrofluid turbulence},\ }\href {https://doi.org/10.1103/PhysRevFluids.9.094604} {\bibfield  {journal} {\bibinfo  {journal} {Phys. Rev. Fluids}\ }\textbf {\bibinfo {volume} {9}},\ \bibinfo {pages} {094604} (\bibinfo {year} {2024})}\BibitemShut {NoStop}%
\bibitem [{\citenamefont {Monin}\ and\ \citenamefont {Yaglom.}(1975)}]{Monin1975}%
  \BibitemOpen
  \bibfield  {author} {\bibinfo {author} {\bibfnamefont {A.~S.}\ \bibnamefont {Monin}}\ and\ \bibinfo {author} {\bibfnamefont {A.~M.}\ \bibnamefont {Yaglom.}},\ }\href@noop {} {\emph {\bibinfo {title} {Statistical Fluid Mechanics: Mechanics of Turbulence}}},\ Vol.~\bibinfo {volume} {2}\ (\bibinfo  {publisher} {MIT Press},\ \bibinfo {year} {1975})\BibitemShut {NoStop}%
\bibitem [{\citenamefont {de~Karman}\ and\ \citenamefont {Howarth}(1938)}]{Karman1938}%
  \BibitemOpen
  \bibfield  {author} {\bibinfo {author} {\bibfnamefont {T.}~\bibnamefont {de~Karman}}\ and\ \bibinfo {author} {\bibfnamefont {L.}~\bibnamefont {Howarth}},\ }\bibfield  {title} {\bibinfo {title} {On the statistical theory of isotropic turbulence},\ }\href {https://doi.org/10.1098/rspa.1938.0013} {\bibfield  {journal} {\bibinfo  {journal} {Proceedings of the Royal Society of London. Series A - Mathematical and Physical Sciences}\ }\textbf {\bibinfo {volume} {164}},\ \bibinfo {pages} {192} (\bibinfo {year} {1938})}\BibitemShut {NoStop}%
\bibitem [{\citenamefont {Frisch}(1995)}]{Frisch1995}%
  \BibitemOpen
  \bibfield  {author} {\bibinfo {author} {\bibfnamefont {U.}~\bibnamefont {Frisch}},\ }\href@noop {} {\emph {\bibinfo {title} {Turbulence: the legacy of A. N. Kolmogorov}}}\ (\bibinfo  {publisher} {Cambridge university press},\ \bibinfo {year} {1995})\BibitemShut {NoStop}%
\bibitem [{\citenamefont {{Kolmogorov}}(1941)}]{Kolmogorov1941}%
  \BibitemOpen
  \bibfield  {author} {\bibinfo {author} {\bibfnamefont {A.}~\bibnamefont {{Kolmogorov}}},\ }\bibfield  {title} {\bibinfo {title} {{The Local Structure of Turbulence in Incompressible Viscous Fluid for Very Large Reynolds' Numbers}},\ }\href@noop {} {\bibfield  {journal} {\bibinfo  {journal} {Akademiia Nauk SSSR Doklady}\ }\textbf {\bibinfo {volume} {30}},\ \bibinfo {pages} {301} (\bibinfo {year} {1941})}\BibitemShut {NoStop}%
\bibitem [{\citenamefont {Argoul}\ \emph {et~al.}(1989)\citenamefont {Argoul}, \citenamefont {Arnéodo}, \citenamefont {Grasseau}, \citenamefont {Gagne}, \citenamefont {Hopfinger},\ and\ \citenamefont {Frisch}}]{Argoul1989}%
  \BibitemOpen
  \bibfield  {author} {\bibinfo {author} {\bibfnamefont {F.}~\bibnamefont {Argoul}}, \bibinfo {author} {\bibfnamefont {A.}~\bibnamefont {Arnéodo}}, \bibinfo {author} {\bibfnamefont {G.}~\bibnamefont {Grasseau}}, \bibinfo {author} {\bibfnamefont {Y.}~\bibnamefont {Gagne}}, \bibinfo {author} {\bibfnamefont {E.~J.}\ \bibnamefont {Hopfinger}},\ and\ \bibinfo {author} {\bibfnamefont {U.}~\bibnamefont {Frisch}},\ }\bibfield  {title} {\bibinfo {title} {Wavelet analysis of turbulence reveals the multifractal nature of the richardson cascade},\ }\href {https://doi.org/10.1038/338051a0} {\bibfield  {journal} {\bibinfo  {journal} {Nature}\ }\textbf {\bibinfo {volume} {338}},\ \bibinfo {pages} {51} (\bibinfo {year} {1989})}\BibitemShut {NoStop}%
\bibitem [{\citenamefont {Banerjee}\ and\ \citenamefont {Galtier}(2014)}]{Banerjee2014}%
  \BibitemOpen
  \bibfield  {author} {\bibinfo {author} {\bibfnamefont {S.}~\bibnamefont {Banerjee}}\ and\ \bibinfo {author} {\bibfnamefont {S.}~\bibnamefont {Galtier}},\ }\bibfield  {title} {\bibinfo {title} {A kolmogorov-like exact relation for compressible polytropic turbulence},\ }\href {https://doi.org/10.1017/jfm.2013.657} {\bibfield  {journal} {\bibinfo  {journal} {Journal of Fluid Mechanics}\ }\textbf {\bibinfo {volume} {742}},\ \bibinfo {pages} {230–242} (\bibinfo {year} {2014})}\BibitemShut {NoStop}%
\bibitem [{\citenamefont {Politano}\ and\ \citenamefont {Pouquet}(1998{\natexlab{a}})}]{Politano1998}%
  \BibitemOpen
  \bibfield  {author} {\bibinfo {author} {\bibfnamefont {H.}~\bibnamefont {Politano}}\ and\ \bibinfo {author} {\bibfnamefont {A.}~\bibnamefont {Pouquet}},\ }\bibfield  {title} {\bibinfo {title} {von k\'arm\'an--howarth equation for magnetohydrodynamics and its consequences on third-order longitudinal structure and correlation functions},\ }\href {https://doi.org/10.1103/PhysRevE.57.R21} {\bibfield  {journal} {\bibinfo  {journal} {Phys. Rev. E}\ }\textbf {\bibinfo {volume} {57}},\ \bibinfo {pages} {R21} (\bibinfo {year} {1998}{\natexlab{a}})}\BibitemShut {NoStop}%
\bibitem [{\citenamefont {Politano}\ and\ \citenamefont {Pouquet}(1998{\natexlab{b}})}]{politano1998dynamical}%
  \BibitemOpen
  \bibfield  {author} {\bibinfo {author} {\bibfnamefont {H.}~\bibnamefont {Politano}}\ and\ \bibinfo {author} {\bibfnamefont {A.}~\bibnamefont {Pouquet}},\ }\bibfield  {title} {\bibinfo {title} {Dynamical length scales for turbulent magnetized flows},\ }\href {https://math.unice.fr/~politano/papers/grl10715.pdf} {\bibfield  {journal} {\bibinfo  {journal} {Geophysical Research Letters}\ }\textbf {\bibinfo {volume} {25}},\ \bibinfo {pages} {273} (\bibinfo {year} {1998}{\natexlab{b}})}\BibitemShut {NoStop}%
\bibitem [{\citenamefont {Galtier}(2008)}]{Galtier2008}%
  \BibitemOpen
  \bibfield  {author} {\bibinfo {author} {\bibfnamefont {S.}~\bibnamefont {Galtier}},\ }\bibfield  {title} {\bibinfo {title} {von k\'arm\'an--howarth equations for hall magnetohydrodynamic flows},\ }\href {https://doi.org/10.1103/PhysRevE.77.015302} {\bibfield  {journal} {\bibinfo  {journal} {Phys. Rev. E}\ }\textbf {\bibinfo {volume} {77}},\ \bibinfo {pages} {015302} (\bibinfo {year} {2008})}\BibitemShut {NoStop}%
\bibitem [{\citenamefont {Pan}\ and\ \citenamefont {Banerjee}(2022)}]{Nandita2022}%
  \BibitemOpen
  \bibfield  {author} {\bibinfo {author} {\bibfnamefont {N.}~\bibnamefont {Pan}}\ and\ \bibinfo {author} {\bibfnamefont {S.}~\bibnamefont {Banerjee}},\ }\bibfield  {title} {\bibinfo {title} {Exact relations for energy transfer in simple and active binary fluid turbulence},\ }\href {https://doi.org/10.1103/PhysRevE.106.025104} {\bibfield  {journal} {\bibinfo  {journal} {Phys. Rev. E}\ }\textbf {\bibinfo {volume} {106}},\ \bibinfo {pages} {025104} (\bibinfo {year} {2022})}\BibitemShut {NoStop}%
\bibitem [{\citenamefont {Galtier}\ and\ \citenamefont {Banerjee}(2011{\natexlab{a}})}]{Galtier2011}%
  \BibitemOpen
  \bibfield  {author} {\bibinfo {author} {\bibfnamefont {S.}~\bibnamefont {Galtier}}\ and\ \bibinfo {author} {\bibfnamefont {S.}~\bibnamefont {Banerjee}},\ }\bibfield  {title} {\bibinfo {title} {Exact relation for correlation functions in compressible isothermal turbulence},\ }\href {https://doi.org/10.1103/PhysRevLett.107.134501} {\bibfield  {journal} {\bibinfo  {journal} {Phys. Rev. Lett.}\ }\textbf {\bibinfo {volume} {107}},\ \bibinfo {pages} {134501} (\bibinfo {year} {2011}{\natexlab{a}})}\BibitemShut {NoStop}%
\bibitem [{\citenamefont {Banerjee}\ and\ \citenamefont {Galtier}(2013)}]{Banerjee2013}%
  \BibitemOpen
  \bibfield  {author} {\bibinfo {author} {\bibfnamefont {S.}~\bibnamefont {Banerjee}}\ and\ \bibinfo {author} {\bibfnamefont {S.}~\bibnamefont {Galtier}},\ }\bibfield  {title} {\bibinfo {title} {Exact relation with two-point correlation functions and phenomenological approach for compressible magnetohydrodynamic turbulence},\ }\href {https://doi.org/10.1103/PhysRevE.87.013019} {\bibfield  {journal} {\bibinfo  {journal} {Phys. Rev. E}\ }\textbf {\bibinfo {volume} {87}},\ \bibinfo {pages} {013019} (\bibinfo {year} {2013})}\BibitemShut {NoStop}%
\bibitem [{\citenamefont {Banerjee}\ and\ \citenamefont {Galtier}(2016{\natexlab{a}})}]{Banerjee2016}%
  \BibitemOpen
  \bibfield  {author} {\bibinfo {author} {\bibfnamefont {S.}~\bibnamefont {Banerjee}}\ and\ \bibinfo {author} {\bibfnamefont {S.}~\bibnamefont {Galtier}},\ }\bibfield  {title} {\bibinfo {title} {An alternative formulation for exact scaling relations in hydrodynamic and magnetohydrodynamic turbulence},\ }\href {https://doi.org/10.1088/1751-8113/50/1/015501} {\bibfield  {journal} {\bibinfo  {journal} {Journal of Physics A: Mathematical and Theoretical}\ }\textbf {\bibinfo {volume} {50}},\ \bibinfo {pages} {015501} (\bibinfo {year} {2016}{\natexlab{a}})}\BibitemShut {NoStop}%
\bibitem [{\citenamefont {Banerjee}\ and\ \citenamefont {Galtier}(2016{\natexlab{b}})}]{Banerjee2016b}%
  \BibitemOpen
  \bibfield  {author} {\bibinfo {author} {\bibfnamefont {S.}~\bibnamefont {Banerjee}}\ and\ \bibinfo {author} {\bibfnamefont {S.}~\bibnamefont {Galtier}},\ }\bibfield  {title} {\bibinfo {title} {Chiral exact relations for helicities in hall magnetohydrodynamic turbulence},\ }\href {https://doi.org/10.1103/PhysRevE.93.033120} {\bibfield  {journal} {\bibinfo  {journal} {Phys. Rev. E}\ }\textbf {\bibinfo {volume} {93}},\ \bibinfo {pages} {033120} (\bibinfo {year} {2016}{\natexlab{b}})}\BibitemShut {NoStop}%
\bibitem [{\citenamefont {Banerjee}\ and\ \citenamefont {Kritsuk}(2017)}]{Banerjee2017}%
  \BibitemOpen
  \bibfield  {author} {\bibinfo {author} {\bibfnamefont {S.}~\bibnamefont {Banerjee}}\ and\ \bibinfo {author} {\bibfnamefont {A.~G.}\ \bibnamefont {Kritsuk}},\ }\bibfield  {title} {\bibinfo {title} {Exact relations for energy transfer in self-gravitating isothermal turbulence},\ }\href {https://doi.org/10.1103/PhysRevE.96.053116} {\bibfield  {journal} {\bibinfo  {journal} {Phys. Rev. E}\ }\textbf {\bibinfo {volume} {96}},\ \bibinfo {pages} {053116} (\bibinfo {year} {2017})}\BibitemShut {NoStop}%
\bibitem [{\citenamefont {Banerjee}\ and\ \citenamefont {Kritsuk}(2018)}]{Banerjee2018}%
  \BibitemOpen
  \bibfield  {author} {\bibinfo {author} {\bibfnamefont {S.}~\bibnamefont {Banerjee}}\ and\ \bibinfo {author} {\bibfnamefont {A.~G.}\ \bibnamefont {Kritsuk}},\ }\bibfield  {title} {\bibinfo {title} {Energy transfer in compressible magnetohydrodynamic turbulence for isothermal self-gravitating fluids},\ }\href {https://doi.org/10.1103/PhysRevE.97.023107} {\bibfield  {journal} {\bibinfo  {journal} {Phys. Rev. E}\ }\textbf {\bibinfo {volume} {97}},\ \bibinfo {pages} {023107} (\bibinfo {year} {2018})}\BibitemShut {NoStop}%
\bibitem [{\citenamefont {Andr\'es}\ and\ \citenamefont {Banerjee}(2019)}]{Nahuel2019}%
  \BibitemOpen
  \bibfield  {author} {\bibinfo {author} {\bibfnamefont {N.}~\bibnamefont {Andr\'es}}\ and\ \bibinfo {author} {\bibfnamefont {S.}~\bibnamefont {Banerjee}},\ }\bibfield  {title} {\bibinfo {title} {Statistics of incompressible hydrodynamic turbulence: An alternative approach},\ }\href {https://doi.org/10.1103/PhysRevFluids.4.024603} {\bibfield  {journal} {\bibinfo  {journal} {Phys. Rev. Fluids}\ }\textbf {\bibinfo {volume} {4}},\ \bibinfo {pages} {024603} (\bibinfo {year} {2019})}\BibitemShut {NoStop}%
\bibitem [{\citenamefont {Felderhof}\ and\ \citenamefont {Kroh}(1999)}]{Felderhof1999}%
  \BibitemOpen
  \bibfield  {author} {\bibinfo {author} {\bibfnamefont {B.~U.}\ \bibnamefont {Felderhof}}\ and\ \bibinfo {author} {\bibfnamefont {H.~J.}\ \bibnamefont {Kroh}},\ }\bibfield  {title} {\bibinfo {title} {Hydrodynamics of magnetic and dielectric fluids in interaction with the electromagnetic field},\ }\href {https://doi.org/10.1063/1.478642} {\bibfield  {journal} {\bibinfo  {journal} {The Journal of Chemical Physics}\ }\textbf {\bibinfo {volume} {110}},\ \bibinfo {pages} {7403} (\bibinfo {year} {1999})}\BibitemShut {NoStop}%
\bibitem [{\citenamefont {Papadopoulos}\ \emph {et~al.}(2012)\citenamefont {Papadopoulos}, \citenamefont {Vafeas},\ and\ \citenamefont {Hatzikonstantinou}}]{Papadopoulos2012}%
  \BibitemOpen
  \bibfield  {author} {\bibinfo {author} {\bibfnamefont {P.~K.}\ \bibnamefont {Papadopoulos}}, \bibinfo {author} {\bibfnamefont {P.}~\bibnamefont {Vafeas}},\ and\ \bibinfo {author} {\bibfnamefont {P.~M.}\ \bibnamefont {Hatzikonstantinou}},\ }\bibfield  {title} {\bibinfo {title} {Ferrofluid pipe flow under the influence of the magnetic field of a cylindrical coil},\ }\href {https://doi.org/10.1063/1.4769177} {\bibfield  {journal} {\bibinfo  {journal} {Physics of Fluids}\ }\textbf {\bibinfo {volume} {24}},\ \bibinfo {pages} {122002} (\bibinfo {year} {2012})}\BibitemShut {NoStop}%
\bibitem [{\citenamefont {Buschmann}(2020)}]{Buschmann2020}%
  \BibitemOpen
  \bibfield  {author} {\bibinfo {author} {\bibfnamefont {M.~H.}\ \bibnamefont {Buschmann}},\ }\bibfield  {title} {\bibinfo {title} {Critical review of heat transfer experiments in ferrohydrodynamic pipe flow utilising ferronanofluids},\ }\href {https://doi.org/https://doi.org/10.1016/j.ijthermalsci.2020.106426} {\bibfield  {journal} {\bibinfo  {journal} {International Journal of Thermal Sciences}\ }\textbf {\bibinfo {volume} {157}},\ \bibinfo {pages} {106426} (\bibinfo {year} {2020})}\BibitemShut {NoStop}%
\bibitem [{\citenamefont {Adrian}\ and\ \citenamefont {Buckmaster}(1981)}]{Adrian1981}%
  \BibitemOpen
  \bibfield  {author} {\bibinfo {author} {\bibfnamefont {R.}~\bibnamefont {Adrian}}\ and\ \bibinfo {author} {\bibfnamefont {J.}~\bibnamefont {Buckmaster}},\ }\bibfield  {title} {\bibinfo {title} {Suppression of turbulence in magnetically stabilized ferroliquids},\ }\href {https://doi.org/https://doi.org/10.1016/0020-7225(81)90031-8} {\bibfield  {journal} {\bibinfo  {journal} {International Journal of Engineering Science}\ }\textbf {\bibinfo {volume} {19}},\ \bibinfo {pages} {303} (\bibinfo {year} {1981})}\BibitemShut {NoStop}%
\bibitem [{\citenamefont {Schumacher}\ \emph {et~al.}(2003)\citenamefont {Schumacher}, \citenamefont {Sellien}, \citenamefont {Knoke}, \citenamefont {Cader},\ and\ \citenamefont {Finlayson}}]{Schumacher2003}%
  \BibitemOpen
  \bibfield  {author} {\bibinfo {author} {\bibfnamefont {K.~R.}\ \bibnamefont {Schumacher}}, \bibinfo {author} {\bibfnamefont {I.}~\bibnamefont {Sellien}}, \bibinfo {author} {\bibfnamefont {G.~S.}\ \bibnamefont {Knoke}}, \bibinfo {author} {\bibfnamefont {T.}~\bibnamefont {Cader}},\ and\ \bibinfo {author} {\bibfnamefont {B.~A.}\ \bibnamefont {Finlayson}},\ }\bibfield  {title} {\bibinfo {title} {Experiment and simulation of laminar and turbulent ferrofluid pipe flow in an oscillating magnetic field},\ }\href {https://doi.org/10.1103/PhysRevE.67.026308} {\bibfield  {journal} {\bibinfo  {journal} {Phys. Rev. E}\ }\textbf {\bibinfo {volume} {67}},\ \bibinfo {pages} {026308} (\bibinfo {year} {2003})}\BibitemShut {NoStop}%
\bibitem [{\citenamefont {Schumacher}\ \emph {et~al.}(2010)\citenamefont {Schumacher}, \citenamefont {Riley},\ and\ \citenamefont {Finlayson}}]{Schumacher2010}%
  \BibitemOpen
  \bibfield  {author} {\bibinfo {author} {\bibfnamefont {K.~R.}\ \bibnamefont {Schumacher}}, \bibinfo {author} {\bibfnamefont {J.~J.}\ \bibnamefont {Riley}},\ and\ \bibinfo {author} {\bibfnamefont {B.~A.}\ \bibnamefont {Finlayson}},\ }\bibfield  {title} {\bibinfo {title} {Effects of an oscillating magnetic field on homogeneous ferrofluid turbulence},\ }\href {https://doi.org/10.1103/PhysRevE.81.016317} {\bibfield  {journal} {\bibinfo  {journal} {Phys. Rev. E}\ }\textbf {\bibinfo {volume} {81}},\ \bibinfo {pages} {016317} (\bibinfo {year} {2010})}\BibitemShut {NoStop}%
\bibitem [{\citenamefont {Schumacher}\ \emph {et~al.}(2011)\citenamefont {Schumacher}, \citenamefont {Riley},\ and\ \citenamefont {Finlayson}}]{Schumacher2011}%
  \BibitemOpen
  \bibfield  {author} {\bibinfo {author} {\bibfnamefont {K.~R.}\ \bibnamefont {Schumacher}}, \bibinfo {author} {\bibfnamefont {J.~J.}\ \bibnamefont {Riley}},\ and\ \bibinfo {author} {\bibfnamefont {B.~A.}\ \bibnamefont {Finlayson}},\ }\bibfield  {title} {\bibinfo {title} {Turbulence in ferrofluids in channel flow with steady and oscillating magnetic fields},\ }\href {https://doi.org/10.1103/PhysRevE.83.016307} {\bibfield  {journal} {\bibinfo  {journal} {Phys. Rev. E}\ }\textbf {\bibinfo {volume} {83}},\ \bibinfo {pages} {016307} (\bibinfo {year} {2011})}\BibitemShut {NoStop}%
\bibitem [{\citenamefont {Altmeyer}\ \emph {et~al.}(2015)\citenamefont {Altmeyer}, \citenamefont {Do},\ and\ \citenamefont {Lai}}]{Altmeyer2015}%
  \BibitemOpen
  \bibfield  {author} {\bibinfo {author} {\bibfnamefont {S.}~\bibnamefont {Altmeyer}}, \bibinfo {author} {\bibfnamefont {Y.}~\bibnamefont {Do}},\ and\ \bibinfo {author} {\bibfnamefont {Y.~C.}\ \bibnamefont {Lai}},\ }\bibfield  {title} {\bibinfo {title} {Transition to turbulence in taylor-couette ferrofluidic flow},\ }\href {https://doi.org/10.1038/srep10781} {\bibfield  {journal} {\bibinfo  {journal} {Scientific Reports}\ }\textbf {\bibinfo {volume} {5}},\ \bibinfo {pages} {10781} (\bibinfo {year} {2015})}\BibitemShut {NoStop}%
\bibitem [{\citenamefont {Mouraya}\ and\ \citenamefont {Banerjee}(2019)}]{Mouraya2019}%
  \BibitemOpen
  \bibfield  {author} {\bibinfo {author} {\bibfnamefont {S.}~\bibnamefont {Mouraya}}\ and\ \bibinfo {author} {\bibfnamefont {S.}~\bibnamefont {Banerjee}},\ }\bibfield  {title} {\bibinfo {title} {Determination of energy flux rate in homogeneous ferrohydrodynamic turbulence using two-point statistics},\ }\href {https://doi.org/10.1103/PhysRevE.100.053105} {\bibfield  {journal} {\bibinfo  {journal} {Phys. Rev. E}\ }\textbf {\bibinfo {volume} {100}},\ \bibinfo {pages} {053105} (\bibinfo {year} {2019})}\BibitemShut {NoStop}%
\bibitem [{\citenamefont {Politano}\ \emph {et~al.}(2003)\citenamefont {Politano}, \citenamefont {Gomez},\ and\ \citenamefont {Pouquet}}]{Politano2003}%
  \BibitemOpen
  \bibfield  {author} {\bibinfo {author} {\bibfnamefont {H.}~\bibnamefont {Politano}}, \bibinfo {author} {\bibfnamefont {T.}~\bibnamefont {Gomez}},\ and\ \bibinfo {author} {\bibfnamefont {A.}~\bibnamefont {Pouquet}},\ }\bibfield  {title} {\bibinfo {title} {von k\'arm\'an--howarth relationship for helical magnetohydrodynamic flows},\ }\href {https://doi.org/10.1103/PhysRevE.68.026315} {\bibfield  {journal} {\bibinfo  {journal} {Phys. Rev. E}\ }\textbf {\bibinfo {volume} {68}},\ \bibinfo {pages} {026315} (\bibinfo {year} {2003})}\BibitemShut {NoStop}%
\bibitem [{\citenamefont {Antonia}\ \emph {et~al.}(1997)\citenamefont {Antonia}, \citenamefont {Ould-Rouis}, \citenamefont {Anselmet},\ and\ \citenamefont {Zhu}}]{Antonia1997}%
  \BibitemOpen
  \bibfield  {author} {\bibinfo {author} {\bibfnamefont {R.~A.}\ \bibnamefont {Antonia}}, \bibinfo {author} {\bibfnamefont {M.}~\bibnamefont {Ould-Rouis}}, \bibinfo {author} {\bibfnamefont {F.}~\bibnamefont {Anselmet}},\ and\ \bibinfo {author} {\bibfnamefont {Y.}~\bibnamefont {Zhu}},\ }\bibfield  {title} {\bibinfo {title} {Analogy between predictions of kolmogorov and yaglom},\ }\href {https://doi.org/10.1017/S0022112096004090} {\bibfield  {journal} {\bibinfo  {journal} {Journal of Fluid Mechanics}\ }\textbf {\bibinfo {volume} {332}},\ \bibinfo {pages} {395–409} (\bibinfo {year} {1997})}\BibitemShut {NoStop}%
\bibitem [{\citenamefont {Podesta}(2008)}]{Podesta2008}%
  \BibitemOpen
  \bibfield  {author} {\bibinfo {author} {\bibfnamefont {J.~J.}\ \bibnamefont {Podesta}},\ }\bibfield  {title} {\bibinfo {title} {Laws for third-order moments in homogeneous anisotropic incompressible magnetohydrodynamic turbulence},\ }\href {https://doi.org/10.1017/S0022112008002280} {\bibfield  {journal} {\bibinfo  {journal} {Journal of Fluid Mechanics}\ }\textbf {\bibinfo {volume} {609}},\ \bibinfo {pages} {171–194} (\bibinfo {year} {2008})}\BibitemShut {NoStop}%
\bibitem [{\citenamefont {Augier}\ \emph {et~al.}(2012)\citenamefont {Augier}, \citenamefont {Galtier},\ and\ \citenamefont {Billant}}]{Augier2012}%
  \BibitemOpen
  \bibfield  {author} {\bibinfo {author} {\bibfnamefont {P.}~\bibnamefont {Augier}}, \bibinfo {author} {\bibfnamefont {S.}~\bibnamefont {Galtier}},\ and\ \bibinfo {author} {\bibfnamefont {P.}~\bibnamefont {Billant}},\ }\bibfield  {title} {\bibinfo {title} {Kolmogorov laws for stratified turbulence},\ }\href {https://doi.org/10.1017/jfm.2012.379} {\bibfield  {journal} {\bibinfo  {journal} {Journal of Fluid Mechanics}\ }\textbf {\bibinfo {volume} {709}},\ \bibinfo {pages} {659–670} (\bibinfo {year} {2012})}\BibitemShut {NoStop}%
\bibitem [{\citenamefont {Galtier}\ and\ \citenamefont {Banerjee}(2011{\natexlab{b}})}]{Banerjee2011}%
  \BibitemOpen
  \bibfield  {author} {\bibinfo {author} {\bibfnamefont {S.}~\bibnamefont {Galtier}}\ and\ \bibinfo {author} {\bibfnamefont {S.}~\bibnamefont {Banerjee}},\ }\bibfield  {title} {\bibinfo {title} {Exact relation for correlation functions in compressible isothermal turbulence},\ }\href {https://doi.org/10.1103/PhysRevLett.107.134501} {\bibfield  {journal} {\bibinfo  {journal} {Phys. Rev. Lett.}\ }\textbf {\bibinfo {volume} {107}},\ \bibinfo {pages} {134501} (\bibinfo {year} {2011}{\natexlab{b}})}\BibitemShut {NoStop}%
\bibitem [{\citenamefont {Banerjee}\ and\ \citenamefont {Andr\'es}(2020)}]{Banerjee2020}%
  \BibitemOpen
  \bibfield  {author} {\bibinfo {author} {\bibfnamefont {S.}~\bibnamefont {Banerjee}}\ and\ \bibinfo {author} {\bibfnamefont {N.}~\bibnamefont {Andr\'es}},\ }\bibfield  {title} {\bibinfo {title} {Scale-to-scale energy transfer rate in compressible two-fluid plasma turbulence},\ }\href {https://doi.org/10.1103/PhysRevE.101.043212} {\bibfield  {journal} {\bibinfo  {journal} {Phys. Rev. E}\ }\textbf {\bibinfo {volume} {101}},\ \bibinfo {pages} {043212} (\bibinfo {year} {2020})}\BibitemShut {NoStop}%
\bibitem [{\citenamefont {Ferrand}(2021)}]{Ferrandthesis}%
  \BibitemOpen
  \bibfield  {author} {\bibinfo {author} {\bibfnamefont {R.}~\bibnamefont {Ferrand}},\ }\href {https://theses.hal.science/tel-03545797} {\emph {\bibinfo {title} {Multi-scale compressible turbulence in astrophysical plasmas viewed through theoretical, numerical and observational methods.}}}\ (\bibinfo  {publisher} {Earth and Planetary Astrophysics, Université Paris-Saclay},\ \bibinfo {year} {2021})\BibitemShut {NoStop}%
\bibitem [{\citenamefont {Taylor}\ \emph {et~al.}(2003)\citenamefont {Taylor}, \citenamefont {Kurien},\ and\ \citenamefont {Eyink}}]{Taylor2003}%
  \BibitemOpen
  \bibfield  {author} {\bibinfo {author} {\bibfnamefont {M.~A.}\ \bibnamefont {Taylor}}, \bibinfo {author} {\bibfnamefont {S.}~\bibnamefont {Kurien}},\ and\ \bibinfo {author} {\bibfnamefont {G.~L.}\ \bibnamefont {Eyink}},\ }\bibfield  {title} {\bibinfo {title} {Recovering isotropic statistics in turbulence simulations: The kolmogorov 4/5th law},\ }\href {https://doi.org/10.1103/PhysRevE.68.026310} {\bibfield  {journal} {\bibinfo  {journal} {Phys. Rev. E}\ }\textbf {\bibinfo {volume} {68}},\ \bibinfo {pages} {026310} (\bibinfo {year} {2003})}\BibitemShut {NoStop}%
\bibitem [{\citenamefont {Mouraya}\ and\ \citenamefont {Banerjee}(2023)}]{Mouraya2023}%
  \BibitemOpen
  \bibfield  {author} {\bibinfo {author} {\bibfnamefont {S.}~\bibnamefont {Mouraya}}\ and\ \bibinfo {author} {\bibfnamefont {S.}~\bibnamefont {Banerjee}},\ }\bibfield  {title} {\bibinfo {title} {{Temperature evolution equation of a compressible turbulent ferrofluid}},\ }\href {https://doi.org/10.1063/5.0128705} {\bibfield  {journal} {\bibinfo  {journal} {Physics of Fluids}\ }\textbf {\bibinfo {volume} {35}},\ \bibinfo {pages} {015120} (\bibinfo {year} {2023})}\BibitemShut {NoStop}%
\bibitem [{\citenamefont {Banerjee}\ \emph {et~al.}(2023)\citenamefont {Banerjee}, \citenamefont {Halder},\ and\ \citenamefont {Pan}}]{Banerjee2023PVNLT}%
  \BibitemOpen
  \bibfield  {author} {\bibinfo {author} {\bibfnamefont {S.}~\bibnamefont {Banerjee}}, \bibinfo {author} {\bibfnamefont {A.}~\bibnamefont {Halder}},\ and\ \bibinfo {author} {\bibfnamefont {N.}~\bibnamefont {Pan}},\ }\bibfield  {title} {\bibinfo {title} {Universal turbulent relaxation of fluids and plasmas by the principle of vanishing nonlinear transfers},\ }\href {https://doi.org/10.1103/PhysRevE.107.L043201} {\bibfield  {journal} {\bibinfo  {journal} {Phys. Rev. E}\ }\textbf {\bibinfo {volume} {107}},\ \bibinfo {pages} {L043201} (\bibinfo {year} {2023})}\BibitemShut {NoStop}%
\bibitem [{\citenamefont {Pan}\ \emph {et~al.}(2024)\citenamefont {Pan}, \citenamefont {Banerjee},\ and\ \citenamefont {Halder}}]{pan2024universal}%
  \BibitemOpen
  \bibfield  {author} {\bibinfo {author} {\bibfnamefont {N.}~\bibnamefont {Pan}}, \bibinfo {author} {\bibfnamefont {S.}~\bibnamefont {Banerjee}},\ and\ \bibinfo {author} {\bibfnamefont {A.}~\bibnamefont {Halder}},\ }\bibfield  {title} {\bibinfo {title} {Universal relaxation of turbulent binary fluids},\ }\href {https://doi.org/10.1038/s42005-023-01498-1} {\bibfield  {journal} {\bibinfo  {journal} {Communications Physics}\ }\textbf {\bibinfo {volume} {7}},\ \bibinfo {pages} {4} (\bibinfo {year} {2024})}\BibitemShut {NoStop}%
\bibitem [{\citenamefont {Banerjee}\ and\ \citenamefont {Halder}(2024)}]{Banerjee2023fundamental}%
  \BibitemOpen
  \bibfield  {author} {\bibinfo {author} {\bibfnamefont {S.}~\bibnamefont {Banerjee}}\ and\ \bibinfo {author} {\bibfnamefont {A.}~\bibnamefont {Halder}},\ }\bibfield  {title} {\bibinfo {title} {{Fundamental units of triadic interactions in Hall magnetohydrodynamic turbulence: How far can we go?}},\ }\href {https://doi.org/10.1063/5.0191882} {\bibfield  {journal} {\bibinfo  {journal} {Physics of Plasmas}\ }\textbf {\bibinfo {volume} {31}},\ \bibinfo {pages} {062301} (\bibinfo {year} {2024})}\BibitemShut {NoStop}%
\bibitem [{\citenamefont {Mortensen}\ and\ \citenamefont {Langtangen}(2016)}]{Mortensen2016}%
  \BibitemOpen
  \bibfield  {author} {\bibinfo {author} {\bibfnamefont {M.}~\bibnamefont {Mortensen}}\ and\ \bibinfo {author} {\bibfnamefont {H.~P.}\ \bibnamefont {Langtangen}},\ }\bibfield  {title} {\bibinfo {title} {High performance python for direct numerical simulations of turbulent flows},\ }\href {https://doi.org/https://doi.org/10.1016/j.cpc.2016.02.005} {\bibfield  {journal} {\bibinfo  {journal} {Computer Physics Communications}\ }\textbf {\bibinfo {volume} {203}},\ \bibinfo {pages} {53} (\bibinfo {year} {2016})}\BibitemShut {NoStop}%
\end{thebibliography}%
	
\end{document}